# Cloud Behaviour on Tidally Locked Rocky Planets from Global High-resolution Modeling

Jun Yang,[1]*† Yixiao Zhang,[1]† Zuntao Fu,[1] Mingyu Yan,[1] Xinyi Song,[1] Mengyu Wei,[1] Jiachen Liu,[1] Feng Ding,[1,2] Zhihong Tan[3]

1. Department of Atmospheric and Oceanic Sciences, School of Physics, Peking University, Beijing 100871, China.
2. Previously at Department of Earth and Planetary Sciences, Harvard University, MA 02138, USA.
3. Program in Atmospheric and Oceanic Sciences, Princeton University, Princeton, NJ 08540, USA.

* Corresponding author: JY, junyang@pku.edu.cn

† These two authors contributed equally to this work.

**This PDF file includes:**
  Main Text (3316 words)
  Figures 1 to 6 (in the main text)
  Methods (4217 words)
  Data Availability
  Code Availability
  Acknowledgements
  Author Contributions
  Competing Interests
  Figure Captions (for the main text figures)
  References (50 in total)

**Supplementary Information in another PDF file:**
  Text (4752 words)
  Tables S1-S4
  Figures S1 to S21
  Videos S1 to S3
  References (27 in total)

**Abstract:** Determining the behaviour of convection and clouds is one of the biggest challenges in our understanding of exoplanetary climates. Given the lack of in situ observations, one of the most preferable approaches is to use cloud-resolving or cloud-permitting models (CPM). Here we present CPM simulations in a quasi-global domain with high spatial resolution (4×4 km grid) and explicit convection to study the cloud regime of 1:1 tidally locked rocky planets orbiting around low-mass stars. We show that the substellar region is covered by deep convective clouds and cloud albedo increases with increasing stellar flux. The CPM produces relatively less cloud liquid water concentration, smaller cloud coverage, lower cloud albedo, and deeper $H_2O$ spectral features than previous general circulation model (GCM) simulations employing empirical convection and cloud parameterizations. Furthermore, cloud streets—long bands of low-level clouds oriented nearly parallel to the direction of the mean boundary-layer winds—appear in the CPM and substantially affect energy balance and surface precipitation at a local level.

**Introduction**

Clouds are critical for planetary climate and habitability since they can absorb and reflect stellar radiation and meanwhile absorb and re-emit thermal infrared radiation[1]. Clouds are also critical for the observational characterizations of exoplanets because they can affect the amplitude of thermal phase curves and can mute the transmission and emission spectral features of atmospheric species[2-5]. The strength of the climatic effects and the degree of the spectral muting depend on the cloud composition, coverage, thickness, altitude, and microphysical properties. Therefore, knowing which types of clouds can form and what characteristics they exhibit are important to understand planetary habitability and the detectability of atmospheric species such as $H_2O$. In this study, for the first time, a global-scale CPM is employed and modified to simulate the convection and clouds and their climatic effects on tidally locked rocky planets orbiting around low-mass stars.

Tidally locked rocky planets are the primary targets for finding potentially habitable planets beyond the solar system. This is due to their frequent transits and large planet-to-star size ratios. Previous studies using global GCMs showed that there are mainly two types of clouds on this kind of planets, deep convective clouds over the substellar region and low-level clouds on the permanent nightside[6-16]. However, the grid sizes of GCMs are hundreds of kilometers, so that small-scale processes such as convection and clouds on scales of meters to kilometers are not resolved and their effects on radiation, momentum, moisture, and energy need to be parameterized. The parameterization schemes involve many empirical equations and parameters based on Earth, raising the question of their applicability to exoplanetary environments that are quite different from Earth.

The System for Atmospheric Modeling (SAM)[17] in a quasi-global domain is used in this study (see Methods and Supplementary Table 1-4 and Supplementary Figs. 1-7). In the model's dynamical core, the hydrostatic approximation is not made and the full vertical momentum equation is solved. The resolution is 4 km in $x$ direction by 4 km in $y$ direction (or 2 km by 2 km) with 48 vertical levels, and the model time step is 10 or 20 s. Two rocky planets, TRAPPIST-1e and K2-72e, are simulated, and both are assumed to be in synchronously rotating orbits. The rotation period of TRAPPIST-1e is 6.1 Earth days, so it is either close to or exactly as the circulation regime of rapidly rotating planets[18-19]. The rotation period of K2-72e is 24.2 days, so it belongs to the slowly rotating regime. Moreover, the effects of cloud albedo feedback on the inner edge of the habitable zone are investigated. The results of SAM are further compared with those from two GCMs, CAM3[20] and ExoCAM[21].

**Results**

**Clouds:** Dense convection clouds form over the substellar area (Fig. 1 and Supplementary Figs. 8-14), and they extend vertically throughout the troposphere. The formation of these clouds is from high convective instability and strong near-surface convergence. Both cloud water amount and cloud fraction on K2-72e are greater than those of TRAPPIST-1e (Fig. 2 & Supplementary Fig. 9). This is due to that the stellar flux of K2-72e is higher than TRAPPIST-1e, 1510 versus 900 W m$^{-2}$, and meanwhile the rotation period is larger, 24.2 versus 6.1 Earth days. In SAM, global-mean planetary albedos are 18% and 37% for TRAPPIST-1e and K2-72e, respectively; these values are smaller than those in the two GCMs. The simulated TRAPPIST-1e's planetary albedos are respectively 29% and 22% in CAM3 and ExoCAM, and the simulated K2-72e's planetary albedos are respectively 47% and 39% in CAM3 and ExoCAM (Supplementary Table 3 & Supplementary Fig. 10A & B). The surface albedo is low (about 3%-7% in global mean), so most of the planetary albedo is contributed by clouds. In longwave radiation, the global-mean cloud effects also have no essential differences, although SAM has relatively smaller values: 11.4, 16.4, and 13.4 W m$^{-2}$ for TRAPPIST-1e and 13.4, 18.6, and 20.6 W m$^{-2}$ for K2-72e in SAM, CAM3, and ExoCAM, respectively (Supplementary Fig. 10C & D).

Comparing SAM to CAM3 and ExoCAM, the simulated cloud water paths are lower in liquid phase but higher in ice phase (Supplementary Fig. 10E-H). This partial compensation between liquid and ice clouds makes the planetary albedo in SAM is only somewhat smaller than those in CAM3 and ExoCAM. Note that the partition between liquid and ice clouds is parameterized based on air temperature in all the three models. In CAM3 and ExoCAM, the fraction of ice cloud mass in total condensate is set to be 100% when air temperature is lower than -40°C and to be 0% when the air temperature is higher than -10°C; between -40°C and -10°C, the fraction of ice cloud mass is a linear function of air temperature[20]. In SAM, a similar piecewise function is used but the two temperature limits[17] are -20°C and 0°C. So, there is more cloud ice water mass in SAM. Sensitivity tests show that the simulated cloud water paths and surface temperature are very sensitive to these two temperature limits employed in the microphysics parameterization scheme of the models (Supplementary Table 3).

Another significant difference is that the simulated cloud fractions in SAM are much lower than those in CAM3 and ExoCAM (Supplementary Fig. 9). In global mean, the simulated cloud coverages in SAM are 26.9% and 41.1% for TRAPPIST-1e and K2-72e, respectively. In contrast, the corresponding values are 94.8% and 81.7% in CAM3 and 90.4% and 97.4% in ExoCAM. The largest difference is at the nightside and at the high latitudes of the dayside. In CAM3 and ExoCAM, clouds are not resolved and cloud coverages are empirically parameterized based on relative humidity, atmospheric stratification, and large-scale updrafts[20]. Because water vapor is trapped in the inversion layer of the nightside, near-surface relative humidity is high there (Supplementary Fig. 15), the parameterized cloud coverages are large in particular for low-level clouds (see also Fig. 1(C) in ref. 6). This bias also occurred when the models were used to simulate Earth's Arctic clouds[22].

Tests of increasing the horizontal resolution from 4 to 2 km and extending the model top from 27 to 48 or 65 km in the quasi-global SAM experiments does not change the main conclusions (Supplementary Figs. 2 & 3). Moreover, small-domain sensitivity tests show that as the horizontal grid spacing in SAM is increased from 6.4 to 1.6, 0.4, or 0.1 km, the change of the substellar clouds is small whereas the nightside cloud water path increases significantly but still less than those in the two GCMs (Fig. 3 & Supplementary Figs. 4 & 10). This might be because the mixing within the boundary layer is the main source of the nightside clouds and a finer resolution can better capture the turbulent motion in small scales. The overall effect of varying the resolution, however, is very weak, within 1.0 W m$^{-2}$ in longwave cloud radiative effect on the nightside, because this layer of clouds is close to the surface and the cloud water path is very small.

Overall, the simulated climates between CPM and GCM are broadly similar, although quantitative differences exist and they are not small in some regional aspects. These experiments confirm that the substellar region of tidally locked habitable planets should be covered by optically-thick clouds especially when the planets have slow rotation rates. The similarity in the results between the two types of models also increases our confidence of using GCMs for the climate simulations of planets in tidally locked orbits, at least for the atmospheric composition employed in this study.

**Stabilizing Cloud Feedback:** Previous studies using GCMs have shown that when stellar flux is increased, more convective clouds form over the substellar region and planetary albedo becomes higher. The clouds act to weaken the original warming caused by the increased stellar flux and to move the inner edge of the habitable zone closer to the host stars, especially for slowly rotating planets[6,8,11,12]. This was called as a 'stabilizing cloud feedback'[6]. In order to more accurately simulate this feedback, we employ SAM to re-investigate this phenomenon and the results are shown in Fig. 4 & Supplementary Fig. 16.

From Fig. 4B, it is clear to see that the planetary albedo increases with increasing stellar flux in all the three models. The increasing rate of the planetary albedo in SAM is close to that in CAM3 but higher than that in ExoCAM. However, the absolute value of SAM's planetary albedo for each stellar flux is only ~1/2 to 3/4 of those in CAM3 and ExoCAM. This means that the background planetary albedo in SAM is relatively lower but its increasing rate is high. The lower albedo in SAM is mainly due to that its simulated cloud liquid water mass is less than those in CAM3 and ExoCAM, although the cloud ice water mass is similar or even higher (Fig. 4C & D).

Among the three models, ExoCAM is the warmest although its planetary albedo is in the middle. This is mainly due to that the greenhouse effect and shortwave absorption by water vapor in ExoCAM are significantly stronger than that in CAM3[9,23,24], because an updated radiative transfer module is employed in ExoCAM[21]. This pushes ExoCAM to enter a runaway greenhouse state at a stellar flux of ~1750 W m$^{-2}$, whereas SAM and CAM3 are still in habitable states even when the stellar flux reaches 2000 W m$^{-2}$. Note that the radiative transfer module used in the SAM experiments is the same as CAM3.

Both the planetary albedo and the stabilizing cloud feedback simulated in SAM are stronger than the previous simulations using a limited-area cloud-resolving model shown in Lefèvre et al. (2021)[25]. In a small domain, 250 km by 250 km at the substellar region, Lefèvre et al. (2021)[25] showed that the planetary albedo is about 6%-10% and its increasing rate is about 3% (absolute value) per an increase of 400 W m$^{-2}$ in the stellar flux. Here, we find that the planetary albedo is about 20% and its increasing rate is about 8% (absolute value) per an increase of 400 W m$^{-2}$ in the stellar flux. One possible reason is that the simulated domain in Lefèvre et al. (2021)[25] was too small to well simulate the strong large-scale near-surface convergence towards the substellar region, which is critical for the formation of large-area dense convective clouds on the dayside.

**Cloud streets:** Between the deep convection clouds over the substellar area and the west/east terminators, the global CPM simulates a special low-level cumulus clouds. Different from the night-side low-level clouds, these clouds exhibit a special structure, characterized by parallel bands of clouds separated by parallel bands of clear-sky air (Fig. 5). The direction of the cloud bands is nearly the same as the mean winds in the planetary boundary layer. These clouds are similar to those observed in convective boundary layer of Earth during cold air outbreaks, which are characterized by cold and dry air flowing fast towards relatively warm neighboring oceans or lands[26-28]. This special cloud distribution looks like straight streets, so that they are called as "cloud streets".

The formation of cloud streets is from the coupling between large-scale circulation and small-scale convection (Supplementary Fig. 17). When the cool air is advected from the nightside to the much warmer dayside, the cool air is heavier and starts to descend, and meanwhile the warm air in the boundary layer is lighter and starts to ascend. Modified by vertical wind shear, wavelike structures develop in the direction nearly parallel to the mean wind[29,30]. In the ascending regions, the air cools, moist condensation occurs, and clouds form, and they are separated by descending regions of clear and relatively dry air. The convection as well as water vapor and clouds is trapped in boundary layer by a strong temperature inversion (Supplementary Fig. 17F), which is maintained by large-scale downwelling and adiabatic compression. Below the inversion, relative humidity is high, ~70-100%, but above the inversion, relative humidity is lower than 30% (Supplementary Fig. 17C).

The characteristics of the cloud streets simulated here are somewhat different from those on Earth. From Fig. 5 and Supplementary Fig. 17, one could find that the spacing of the cloud bands ranges from ~10 to 100 km, generally larger than that on Earth, ~2 to 20 km[30]. The convection depths are similar, ~2 km. These mean that the aspect ratios (i.e., cloud street wavelength divided by the convection depth) are ~5-50, greater than the values of 2 to 20 with the most frequent values being ~3 to 4 on Earth[30]. This may be due to the fact that the vertical shear of the horizontal winds near the top of the boundary layer is stronger than that on Earth (Supplementary Fig. 17E). The larger horizontal winds are due to the dramatic horizontal day-night surface temperature contrast on the tidally locked planets, ~30-70 K (Supplementary Figs. 8 & 16). Under a stronger vertical wind shear, the absolute value of the Richardson number ($R_i$, an index of instability) is smaller, thereby the maximum growth rate of an unstable perturbation occurs at a smaller wavenumber, i.e., at a greater wavelength[31].

The cloud streets have significant effects on energy balance and surface precipitation, but only in local regions, as shown in Supplementary Fig. 18. In the updraft regions of the cloud streets, there is more cloud water and cloud top temperatures are lower, so that planetary albedo is higher, downward shortwave radiation reaching the surface is smaller, outgoing longwave radiation to space is smaller, and surface precipitation is larger, compared to the downdraft regions between the cloud streets. The clear-sky regions between updrafts allow the atmosphere to emit thermal radiation to space more easily than that in an alternative condition when all the regions were covered by uniform clouds. In this view, the clear-sky downdraft regions act like a number of small 'radiator fins'. This is similar to the effect of the large radiator fin of the dry subtropics on Earth, which acts to stabilize the tropical climate[32]. In area, the cloud street region covers about 10% to 30% of the dayside (Fig. 1), which indicates that the cloud streets can have significant local climatic effects, although less important than the substellar deep convective clouds.

Cloud streets were not found in previous GCM simulations for tidally locked planets. In GCMs[6,9], the coupling between large-scale circulation and grid-scale convection was considered, but horizontal resolutions (hundreds of kilometers) were too coarse to resolve the cloud streets, and GCMs' convection processes were empirically parameterized. Similar patterns to the cloud streets could be vaguely seen in previous high-resolution simulations such as Fig. 1 of Zhang et al. (2017)[33] and Fig. 12 of Sergeev et al. (2020)[34]. But, neither of them had realized that these clouds belong to cloud streets, and the structure, underlying mechanism, and climatic effects were not analyzed. For more details on the convection and precipitation, please see supplementary Figs. 19-21 and Supporting Information.

**Observational Characteristics:** The different concentrations and spatial patterns of clouds simulated by the models can influence the projection of observational characteristics, as shown in Fig. 6. For TRAPPIST-1e, phase curves obtained from the three models are similar in shape, with the dayside infrared emission being higher than the nightside, generally following the distribution of surface temperature. The peak of each phase curve exhibits a westward shift related to the substellar point (Fig.

6A). This westward shift is due to that equatorial superrotation transports clouds towards the east side of the substellar point and the infrared emission from the cloud top is small because of the low cloud top temperatures. The degree of the westward shift is ~5º, 20º, and 40º in CAM3, ExoCAM, and SAM, respectively.

For K2-72e, all the three models show a hump-like shape in the thermal phase curves with a minimum value at the orbital phase angle between -20º and 90º and one maximum value on each side (Fig. 6B). ExoCAM shows a minimum at the orbit phase angle of about 0º (viewing the dayside). This is due to that liquid cloud water path in ExoCAM is much higher than those in SAM and CAM3 (Supplementary Fig. 10F & Table 3), and the clouds absorb more thermal emission from the surface and meanwhile the cloud-top temperature is relatively lower.

Figures 6C & E show the transmission spectra of TRAPPIST-1e for the morning terminator and the evening terminator, respectively. For the evening terminator, all the three models exhibit very weak signals, less than 2-3 ppm (parts per million) at the 1.4 μm wavelength of $H_2O$ molecular absorption features (which does not overlap with $CO_2$ absorption). This is due to that dense clouds cover the evening terminator (Fig. 2; see also references 3 & 35). For the morning terminator, the atmospheric signal at 1.4 μm simulated in SAM, ~6-10 ppm, is deeper than those in CAM3 and ExoCAM and also the model LMDG used in the study of Fauchez et al. (2019)[2], ~2-3 ppm. It is due to the fact that the morning terminator of SAM has less clouds (Fig. 2A).

Figures 6D & F show the transmission spectra of K2-72e. Its transmission spectral features are shallower than those for TRAPPIST-1e, such as ~3 versus ~10 ppm at 1.4 μm for the morning terminator. This is due to two factors: 1) the star K2-72 is much larger than TRAPPIST-1, 0.33 versus 0.12 of Sun's radius, and the transit depth is inversely proportional to the stellar projected area; and 2) the atmospheric scale height of K2-72e is ~78% of TRAPPIST-1e, due to the larger gravity (1.29 versus 0.93 Earth's gravity), although K2-72e is ~20 K warmer than TRAPPIST-1e. Moreover, K2-72e is ~217 light years away from Earth and TRAPPIST-1e is ~40 light years away, which influences the signal-to-noise ratio (SNR) of observations. Therefore, the detection of $H_2O$ molecule on K2-72e would be harder than TRAPPIST-1e.

**Discussions**

In this study, for the first time, a quasi-global CPM is employed to simulate the clouds and climate of tidally locked habitable planets having Earth-like atmospheric compositions. These simulations show dense convective clouds over the substellar region, a stabilizing cloud feedback as increasing stellar flux, and a special type of clouds—cloud street. For time- and global-mean, the simulated surface temperature and planetary albedo in the CPM are broadly similar to those found in GCM experiments, but GCM simulations may have overestimated the cloud liquid water amount and cloud coverage especially on the nightside, and have also overestimated the effect of clouds on the transmission spectra of $H_2O$ molecule.

This work improves our understanding of convection and clouds and the interactions between convection, cloud, and circulation on tidally locked rocky planets. It also opens the door of global-scale cloud-resolving simulations for exoplanets. Our results may be used to benchmark and improve convection and cloud parameterizations in GCMs, an important subject for future exoplanet model development. Further work is required to use high-resolution models to examine the inner edge (higher insolation) and the outer edge (denser $CO_2$) of the habitable zone for both locked and non-locked planets. This would have a number of implications for designing telescopes and for finding habitable exoplanets.

Finally, we emphasize that even in high-resolution cloud simulations, there are considerable uncertainties. One is from microscale processes, such as ice aggregation rate, evaporation of rain droplets, and the partitions of cloud water, cloud ice, and rain[17,37]. These processes are still required to be parameterized and the parameterization scheme is not unique. Another one is from model resolution. For perfect simulations of stratus and stratocumulus clouds, a minimum horizontal grid spacing of $\mathcal{O}(100)$ m and a minimum vertical grid resolution of $\mathcal{O}(10)$ m are required[38]. For large-domain simulations, these requirements are far beyond present computation powers.

**Methods**

**Exoplanet TRAPPIST-1e and K2-72e**: Two confirmed terrestrial planets, TRAPPIST-1e[39] and K2-72e[40], are simulated in this study. The planetary parameters are listed in Supplementary Tables 1 & 2. TRAPPIST-1e is somewhat smaller than Earth (0.91 Earth's radius) and its orbital period (= rotation period) is ~6.1 Earth days, and K2-72e is larger than Earth (1.29 Earth's radius) and its orbital period (= rotation period) is ~24.2 Earth days. Stellar spectra of ~2500 and ~3400 K are used for TRAPPIST-1e and K2-72e, respectively. The atmosphere is assumed as earth-like, $10^4$ kg m$^{-2}$ $N_2$ plus 355 ppmv $CO_2$ and variable $H_2O$. There are no other greenhouse gases, aerosols, $O_2$, or $O_3$. The surface gravities of the two planets are 0.93 and 1.29 of the Earth's value (9.81 m s$^{-2}$), respectively. Due to the different gravities, the mean surface pressure is ~0.93 bar for TAPPIST-1e but ~1.29 bar for K2-72e.

The atmospheric circulation of TRAPPIST-1e is in or close to the regime of rapid rotation[19], while K2-72e is in the slow rotation regime. Another interesting, slowly rotating planet is TOI-700d, which has a rotation period of ~37.4 Earth days[41]. The effective temperatures of the host stars are similar, ~3480 and ~3360 K for TOI-700 and K2-72, respectively. We choose K2-72e rather than TOI-700d, because the stellar flux receiving on the planet of K2-72e is greater, ~1510 vs ~1183 W m$^{-2}$, although TOI-700d is somewhat closer to Earth, ~101.4 vs ~217.1 light years away. The greater stellar flux can trigger more convective clouds[6,9]; this make us easier to compare the differences between SAM and GCMs.

**Cloud-permitting simulations:** The model we employ here is the System for Atmospheric Modeling (SAM)[17] version 6.11.6. The model is based on the anelastic dynamical equations with bulk microphysics and without cumulus parameterization. Prognostic quantities in the model are the three velocity components, liquid/ice water static energy, non-precipitating water (water vapor, cloud water, and cloud ice), and precipitating water (rain, snow, and graupel). Sub-grid momentum, moisture, and energy fluxes are evaluated using a Smagorinsky-type prognostic closure. Cloud microphysical processes on scales of microns to millimeters, such as evaporation and sublimation in the atmosphere and precipitation formation via collisions, are still required to be parameterized. In this study, the experiments use the computationally efficient single-moment microphysics scheme.

The model uses a Cartesian geometry rather than a spherical geometry, because the spherical geometry is not yet allowed in SAM. The horizontal resolution is 4 km in latitude by 4 km in longitude or 2 km by 2 km (Supplementary Table 2). The computational domain is in a global range extending from 90°S to 90°N in latitude and from 0° to 360° in longitude but in the Cartesian geometry. So, we call the experiments as 'quasi-global cloud-permitting', rather than 'global cloud-resolving' over a sphere or 'near-global cloud-resolving' in a Cartesian geometry from 46°S to 46°N (used in such as Bretherton & Khairoutdinov (2015)[42]). Due to this, atmospheric circulation in polar regions simulated in this frame is not credible. However, to facilitate view of the results, the zonal and meridional coordinates of the figures shown in this paper are given in degrees of longitude and latitude, respectively. The nightside covers the longitudes smaller than 90° or larger than 270°, where the stellar flux is zero. Due to the different planetary radii, one degree of longitude is equal to ~103 km for TRAPPIST-1e but ~143 km for K2-72e, and the same for one degree of latitude. For the limitations of the Cartesian geometry, please see the discussions in Supporting Information.

A latitudinally-dependent Coriolis parameter is used in the simulations, f = 2Ωsinφ, where Ω is the planetary rotation rate and φ is the latitude. So, the Coriolis force (as well as the beta effect) is included. For TRAPPIST-1e, the planetary rotation rate is 0.16 of modern Earth, and for K2-72e, it is 0.04 of modern Earth. The Coriolis parameter is a key factor in the climate simulations[43].

**Two steps for each experiment**: There are two steps for each experiment. Firstly, a quasi-global experiment with a 40-km resolution was run, and it was initialized with no wind and horizontally uniform surface temperature (300 K) and saturated humidity profiles. The surface is coupled to a slab ocean with a constant depth of 1.0 m. Each 40-km experiment was run for 210 Earth days, by which time it has reached an approximate statistical equilibrium state (Supplementary Fig. 1). This spin-up procedure is able to greatly reduce the required computation time for the next step. Note that the ocean depth is 1 m in the 40-km experiments. Because there is neither seasonal nor diurnal cycle on the synchronously rotating planets, so the depth of the slab ocean has a very small influence on the climate. In the sensitivity test of Yang et al. (2013)[6], changing the slab ocean depth from 1 m to 50 m affects the global-mean surface temperature by ~1 K (see their online Table 1). Secondly, the instantaneous variables (including air temperature, winds, water vapor, clouds, etc.) of the final quasi-equilibrium fields of the 40-km experiment is linearly interpolated to a 4-km grid in also a quasi-global domain. Then, we run the model for 30 or 50 Earth days. In the 4 km experiment, the sea surface temperatures (SSTs) are fixed to the time-mean values of the last 30 Earth days of the 40-km experiment. The time step of the experiments is 10 or 20 s, but radiative fluxes and heating rates are updated every 15 mins. The setup of these two steps is similar to that used in Bretherton & Khiroutdinov (2015)[42].

Due to computation resource limits, the 4 km experiments did not use a slab ocean or run for a longer time than 50 Earth days. The results are shown in Figs. 1-2 and Supplementary Figs. 1-2 and online Supplementary Video 1-3. We have added one 2 km experiment but also with fixed SSTs (from the 4-km experiment) and several regional experiments (described below) with higher resolutions. All these experiments showed that the main conclusions are likely robust except that the night-side clouds strongly depend on the horizontal resolution (see Fig. 3 & Supplementary Fig. 4).

In the experiments, the timescale of convection is less than hours, the timescale of the global atmospheric overturning circulation is about 30-100 Earth days, and the radiative timescale is about 100 Earth days. The required model time to reach equilibrium is mainly determined by the greatest one, the radiative timescale. So, the simulation length of 50 Earth days (under 4 km resolution) is not long enough to let the system reach perfect statistical equilibrium. But, due to the antecedent 210 days run under the 40 km resolution in the first step, it requires a shorter time to reach quasi-equilibrium for the 4 km experiments. In Bretherton & Khiroutdinov (2015)[42], they stated that "The spin-up on the 4 km grid mainly just fills in small-scale variance unresolved by the 20 km (*40 km in this study*) grid, with some adjustments in statistics of cloud and humidity-related fields on larger scales." Therefore, the short run of 50 Earth days is roughly suitable for the target in this study; of course, further longer experiments would be better to examine the time-mean and transient characteristics of the system.

Another time scale is related to the thermal inertia of the ocean and atmosphere. For a well-mixed slab ocean, the strength of its thermal inertia is determined by $\rho_o C_p^o H$, where $\rho_o$ is seawater density, $C_p^o$ is the specific heat capacity of seawater, and $H$ is the slab ocean depth. For the atmosphere, the strength of its thermal inertia is approximately determined by $\rho_o C_p^o H_{eq}$, where $H_{eq}$ is the equivalent water mixed layer depth of the atmosphere. The value of $H_{eq}$ can be calculated as $P_s/(g\rho_o) \times C_p^a/C_p^o$, where $P_s$ is the mean surface pressure, $g$ is the planetary surface gravity, and $C_p^a$ is the specific heat capacity of the atmosphere. For Earth-like atmosphere, $P_s$ is ~1.0 bar, $g$ is ~10 m s$^{-2}$, and $C_p^a/C_p^o$ is approximately equal to 0.24, which is 1004/4218. So, the equivalent water depth of the atmosphere is about 2.4 m. If slab ocean depth is 50 m, the equivalent mixed layer depth for the thermal inertia of the atmosphere-ocean system is roughly 52.4 m. This means that in one experiment with the slab ocean depth of 50 m, the timescale for the system to reach equilibrium is about 21.8 (= 52.4/2.4) times that of one experiment with fixed SST. If slab ocean depth is 1 m, the equivalent mixed layer depth of the atmosphere-ocean system decreases to 3.4 m. This means that in one experiment with the slab ocean depth of 1 m, the timescale for the system to reach equilibrium is about 1.4 (= 3.4/2.4) times that of one experiment with fixed SST. If further consider that the atmosphere and ocean are not perfectly in synchronous responses to an external forcing, the two ratios (21.8 and 1.4) would likely be larger. These analyses suggest that the equilibrium time scale with a very shallow slab ocean depth does not differ significantly from that of a fixed-SST experiment. Therefore, for improving the simulations in the future, a slab ocean should be employed instead of fixed SST, provided there are sufficient computational resources.

**Vertical resolution and the top of the model:** The model has 48 vertical levels from the surface to about 27 km with variable grid spacing, which is suitable to resolve the troposphere. Boundary conditions are periodic in the zonal direction and free-slip rigid walls at northern and southern boundaries; there is no cross-pole mass or energy transport in SAM. A sponge layer in the upper 9 km of the model is included in order to minimize the reflection of gravity waves from the model top.

In order to test the effects of varying the vertical resolution and model top, we do several additional experiments with increasing the number of the vertical levels from 48 to 72 and meanwhile raising the top of the model from 27 to 42 or 65 km. The results are shown in the Supplementary Fig. 3. Overall, these two parameters do not influence the conclusions of this study, although they do affect some details.

**Small-domain high-resolution experiments**: We set up a series of small-domain simulations which use several different grid sizes to further examine the robustness of our experiments. These simulations use the same version of SAM as the quasi-global simulations but run in rectangular domains with periodic boundary conditions on both zonal and meridional sides and are forced by the large-scale temperature, wind, and moisture profiles from the quasi-global 4 km simulations at certain locations. We choose the sub-stellar point (SP) and the anti-stellar point (AP) as the locations for the small-domain simulations.

We have done four groups of simulations, two for TRAPPIST-1e and another two for K2-72e. Each group consists of four simulations with horizontal grid sizes of 6.4, 1.6, 0.4, and 0.1 km. The number of grid points in the small domain (Nx × Ny × Nz) is 32 × 32 × 48, and it is the same for all the small-domain experiments. For all these simulations, the vertical grid is identical to that of quasi-global simulations. For the SP experiments, the incident stellar radiation is uniform in the domain, the stellar strength and the incidence angle are the same as the corresponding location in quasi-global experiments. For the AP experiments, these is no incident stellar radiation. The Coriolis force is not considered, since both SP and AP are located on the equator.

To represent the large-scale forcing from the quasi-global simulation, we apply 1) the nudging of horizontal velocity, temperature, and specific humidity with a relaxation timescale of two hours, and 2) a large-scale vertical velocity forcing from the reference profile. The total large-scale forcing tendency can be expressed as

$$\left(\frac{\partial \lambda}{\partial t}\right)_{l.s.} = -\frac{\lambda - \lambda_{\text{ref}}}{\tau} - w_{\text{ref}} \frac{\partial \lambda}{\partial z}$$

where $\lambda$ represents the liquid-ice static energy, the specific humidity, or the wind components in the x or y directions; $\lambda_{\text{ref}}$ is the corresponding reference profile; specifically, $w_{\text{ref}}$ is the reference vertical velocity; and $\tau$ is equal to two hours. The reference profile is the timely and horizontally averaged profile from the quasi-global simulation in a square with a side length of 160 km, the center of which is the SP or AP. We use the data between Days 230 to 239 for time averaging. Although it is possible to use reference profiles that are varying with time, we use a fixed time-mean profiles for simplification and for noise reduction. The surface temperature (uniform) is also fixed to the mean value of the same region. This method is similar to that used in references 44 & 45.

We run each simulation for 30 days starting from the corresponding reference profiles. Figure S4 shows that the total cloud water content reaches equilibrium in no more than 10 Earth days in all the simulations. For the two SP simulations, the total cloud water path is insensitive to the grid size with an overall relative error of approximately 10%. These values are consistent with our quasi-global simulations (see also Fig. 3). These results suggest that a cloud-permitting resolution of 4 km for the dayside can possibly produce almost the same results of a resolution as fine as 0.1 km. For the AP simulations, a coarse resolution results in an underestimated cloud water path (Fig. 3 & Supplementary Fig. 4).

**Stabilizing cloud feedback experiments:** Besides the simulations of the two planets TRAPPIST-1e and K2-72e, a series of experiments has been performed to examine the response of the tidally locked climate to the orbit of being closer and closer to the host star, using the three models (SAM, CAM3, and ExoCAM). Six experiments are performed for each model. The star temperature is 3300 K, and planetary radius and gravity are the same as Earth's values. Atmospheric composition is $N_2$ of $10^4$ kg m$^{-2}$ with variable water vapor; there is no $CO_2$, $N_2O$, $CH_4$, $O_2$, or $O_3$. These experiments are useful in knowing the possible effect of cloud feedback on the location for the inner edge of the habitable zone. Note that the values of greenhouse gas concentrations (such as $CO_2$) can influence the exact level of surface temperature but cannot affect the trend under increasing stellar flux. For example, Yang et al. (2013)[6] showed that increasing $CO_2$ concentration from zero to 355 ppmv causes an increase of global-mean surface temperature by 7.5 K (see the online version of their full Table 1).

For SAM, a lower resolution (than 4 km by 4 km) is used in this group of experiments, due to computation resource limitations. Again, for each experiment, there are two steps. The first step is using a relatively low resolution of 40 km by 40 km, and the second step is using a relatively higher resolution of 20 km by 20 km. The initial profiles of the second step are from the quasi-equilibrium fields of the first step through linear interpolation. So, these experiments could be called as cloud-permitting simulations, rather than cloud-resolving simulations. In both steps, the atmosphere is coupled to a 1-m slab ocean. The time series of the results are shown in Supplementary Fig. 5. Each experiment was run for 350 Earth days in total; all the experiments have already reached quasi-equilibrium. For CAM3 and ExoCAM, the surface is coupled to a slab ocean of 50 m everywhere, and each experiment is run for ~60 Earth years and the last 5 years are used for analyses. The main results are showed in Fig. 4 and Supplementary Fig. 16. The behavior of convection and precipitation under extremely hot climates are discussed in the Supporting Information and showed in Supplementary Figs. 19 & 20.

Note that the treatment of cloud particle sizes is the same for the three models, which makes the comparisons between the three models be a little more straightforward. The effective droplet radius for liquid water clouds is specified to be 14 μm over ocean. For ice water clouds, the effective droplet radius is a non-monotonic function of air temperature[20]. But, the partition between liquid and ice clouds are different between SAM and CAM3/ExoCAM; see Supporting Information for discussing this.

**Testing the method of quasi-global cloud-permitting simulation:** In recent years, the method described above has been successfully used in the simulations of clouds, circulation, and their interactions on an aqua-Earth[42,46]. We have also done two experiments of a global-scale aqua-Earth. Its domain is a zonally periodic 20000 km-long equator-centered channel with latitudinally varying surface temperatures and spans from 60°S to 60°N (the first experiment) or 90°S to 90°N (the second experiment). The horizontal resolution is 10.4 km in longitude by 13.9 km in latitude. The surface is covered by ocean everywhere. Surface temperatures of $Q_{OBS}$ that used in the Aqua-Planet Experiment project[47] are specified in the simulation: It is zonally and hemi-spherically symmetric but decreases as a function of latitude from 300 K at the equator to ~273 K at 60°S and 60°N. In the region between 60°S-90°S and 60°N-90°N of the second experiment, the surface temperature is a constant, 273 K. There is no polar sea ice or snow in these two experiments. Surface gravity is the same as Earth's value, and the mean surface air pressure is ~1.0 bar ($N_2$) with a $CO_2$ concentration of 369 ppmv. The rotation period is one Earth day, and a latitude-dependent Coriolis parameter and a Cartesian geometry are used. The simulation results (Supplementary Figs. 6 & 7) and some discussions are shown in the Supporting Information. Overall, the results suggest that the method used here is suitable, although not perfect.

**Global GCM simulations:** In order to compare the results of SAM with other models, two AGCMs (CAM3 and ExoCAM) are employed to run for the corresponding experiments of TRAPPIST-1e and K2-72e. CAM3 solves the primitive equations over a rotating sphere, developed at the National Center for Atmospheric Research of USA[20]. The radiative transfer module used in CAM3 is the same as that in SAM. Clouds, convection, condensation, precipitation, and boundary layer mixing are parameterized in the model. The horizontal resolution is 3.75º in latitude and 3.75º in longitude with 26 vertical levels from near surface to the level of ~3 hPa. The surface is coupled to a slab ocean of 50 m deep everywhere. Sensitivity tests with a slab ocean depth of 1.0 m show that the ocean depth does not influence the equilibrium surface temperatures; this is due to the fact that there is neither diurnal nor seasonal cycle in the experiments. All other parameters such as planetary orbits are the same as those in the SAM experiments. ExoCAM is similar to CAM3, except that its radiative transfer module is more accurate, and it is able to well simulate moist and runaway greenhouse states, which are directly linked to the inner edge of the habitable zone[21]. For planets in the middle range of the habitable zone, ExoCAM is similar to CAM3[24]. ExoCAM was run under a horizontal resolution of 4º in latitude by 5º in longitude. Each experiment was run for 60 Earth years and the last 5 years are used for analyses. The results are showed in Figs. 2 & 4 and Supplementary Figs. 8-14.

**Radiative transfer modules:** In SAM, longwave and shortwave radiative heating rates are calculated interactively using the CAM radiation scheme[20]. The radiation transfer is roughly suitable as long as surface temperature is less than ~320 K and $CO_2$ concentration is less than ~$10^5$ ppmv, and there are moderate differences in $H_2O$ and $CO_2$ radiative forcing between CAM and line-by-line radiative transfer models[23,48]. So, SAM is able to well simulate the climate of planets in the middle range of the habitable zone, but not for planets near the inner or outer edge of the habitable zone. ExoCAM has an update, more accurate radiative transfer scheme[21,9]. The differences in the radiative transfer scheme between CAM3 (also used in SAM) and ExoRT (used in ExoCAM) were addressed in references 23, 24, & 9. ExoRT has stronger greenhouse effect and larger shortwave absorption by water vapor, as shown in Figs. 3 & 7 in Yang et al. (2016)[23]. When the surface temperature is below 300 K, differences between the two modules are small. At higher temperatures, the differences are large: In longwave radiation, the difference in outgoing longwave radiation can reach ~20 W m$^{-2}$, and in shortwave radiation, the difference in absorbed shortwave energy by water vapor can reach tens of W m$^{-2}$. If SAM's radiative transfer module is replaced with ExoRT, the surface will become much warmer, as suggested in the simulations of Kopparapu et al. (2017, see their[9] Figure 2).

In all the experiments, the surface is covered by ocean everywhere, or called 'aqua-planet'. No sea ice is included because sea ice module has not yet been incorporated in SAM. Local surface albedo is between ~2% and ~30%, depending on the solar zenith angle and the stellar spectrum. The solar zenith angle is a function of both latitude and longitude, but it has neither seasonal nor diurnal cycle due to the permanent day and night of the simulated planets.

**Thermal phase curve and transmission spectra calculations:** Thermal phase curve is temporal variations of the disk-integrated broadband infrared emission of the planet as a function of its orbital phase angle, observed by a distant observer[49]. It is mainly determined by surface/air temperature, atmospheric composition, cloud/haze, and horizontal energy transport. Transmission spectrum is the apparent radius of the planet as a function of wavelength, measured when the stellar light travels through the planetary limbs[4]. It is mainly determined by stellar spectrum, atmospheric composition, atmospheric scale height, and cloud/haze at the terminators (i.e., the longitudes around 90º and 270º as shown in Fig. 2).

The Planetary Spectrum Generator (PSG) developed by NASA[50] is used in this study. Firstly, profiles of air temperature, pressure, water vapor concentrations, cloud mixing ratios, and cloud particle sizes of each latitude obtained from the climate models are used as the input data of PSG to calculate the transit spectrum. Then, the average of all the latitudes is calculated for the east terminator and for the west terminator, following Song & Yang (2021)[35]. A transit spectrum of the full disk-integrated atmosphere (if required) could be simply the average between the east-terminator spectrum and the west-terminator spectrum. The relative transit depth is approximately equal to $2R_p \delta R / R_s^2$, where $R_p$ is the planetary radius of the solid surface, $\delta R$ is the transit atmospheric thickness in altitude, and $R_s$ is the radius of the host star. For TRAPPIST-1 and K2-72, the star's radius is $0.84 \times 10^8$ km and $2.30 \times 10^8$ km, respectively. For each planet, we calculate the spectrum over 30 instantaneous moments and then do the average.

For transmission spectrum, wavelengths between 0.6 and 5.0 $\mu$m are calculated under a resolving power of 300. These wavelengths are the best choice for the atmospheric characterizations of terrestrial planets due to relatively small instrumental noise expectations[36]. Only the range between 0.6 and 1.7 $\mu$m is shown in Fig. 6 because this range is the best for detecting $H_2O$ molecule in the atmosphere. The 1.4 μm feature does not overlap with a $CO_2$ feature. Besides, the wavelengths[5] around 6 μm are also appropriate for detecting $H_2O$. There are some overlaps between $CO_2$ and $H_2O$ at ~2.0 and 2.7 $\mu$m and between $CO_2$ and $N_2$-$N_2$ at 4.3 $\mu$m. Moreover, previous studies had already clearly shown that detecting $CO_2$ on tidally locked habitable planets using *JWST* is possible[2,3,4,36], so in this study we focus on the detectability of $H_2O$ only. For the temporal variability of the transmission spectra, please see refs. 3 & 36.

By default, we did not extrapolate the model top to a lower pressure in the calculations. In order to know its effect, we have done one test by extrapolating the model top to 0.1 Pa, using the method of "Intermediate" proposed in Suissa et al. (2020)[5]. We find that the effect of extrapolating the model top is small, within 1 ppm. This result is consistent with the finding in Suissa et al. (2020)[5]: For a non-runaway planet, additional layers will not have a great effect, as long as the original model top is higher than the cloud deck.

**Additional Information**: Supplementary information is available for this paper at xxx.

**Correspondence and requests for materials** should be addressed to JY, Email: junyang@pku.edu.cn.

**Figure captions:**

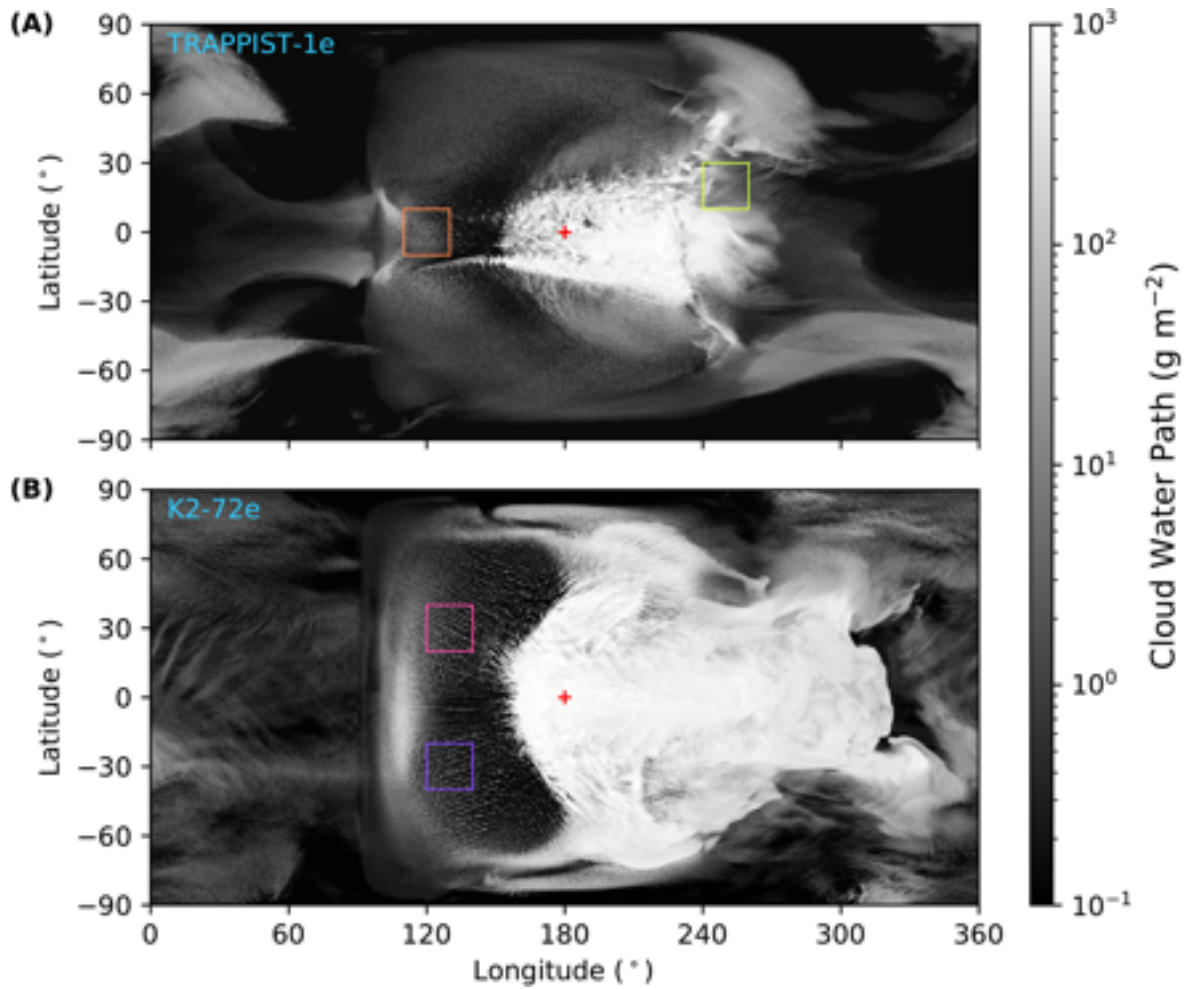

**Figure 1. Maps of instantaneous vertically-integrated cloud liquid and ice water amount in the SAM experiments with a horizontal resolution of 4 km.** Panel (A) is TRAPPIST-1e, and panel (B) is for K2-72e. The red cross marks the substellar point. This figure is the same as Fig. 2A & B below but for transient snapshots rather than time-mean maps. For the time series of cloud water path maps, please see Supplementary Videos 1-3 online. The small squares with boundary colors of yellow, green, pink, and blue are selected regions for showing local cloud streets in Fig. 5 below.

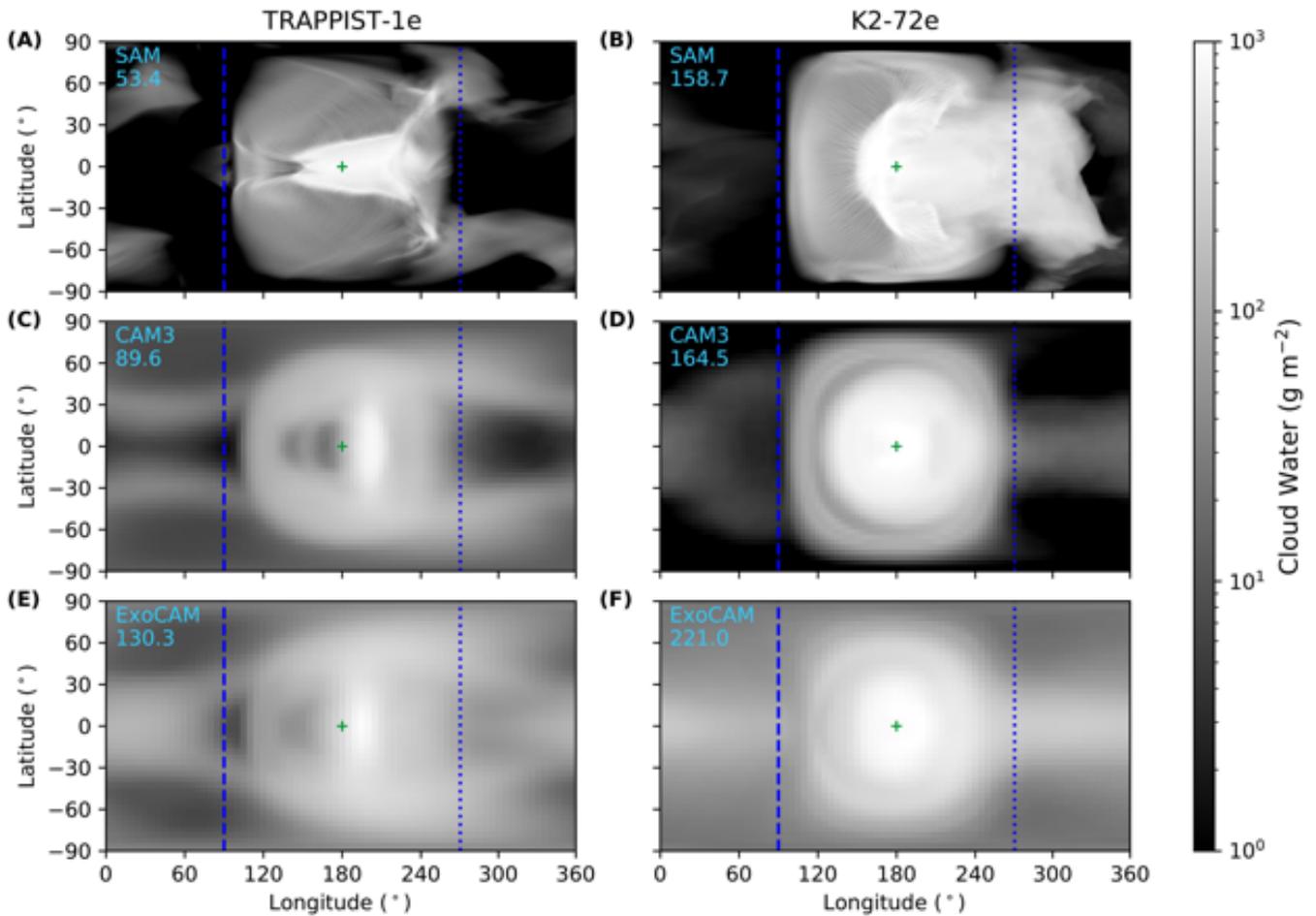

**Figure 2. Time-mean vertically-integrated cloud water amount (including both liquid and ice) simulated by the three models.** Left panels are for TRAPPIST-1e, and right panels are for K2-72e. The global-mean values are listed in the top left corner of each panel. The red cross marks the substellar point. The blue dashed line is the morning terminator and the blue dotted line is the evening terminator, used for transmission spectra calculations (Fig. 6 below). SAM has less total cloud water path.

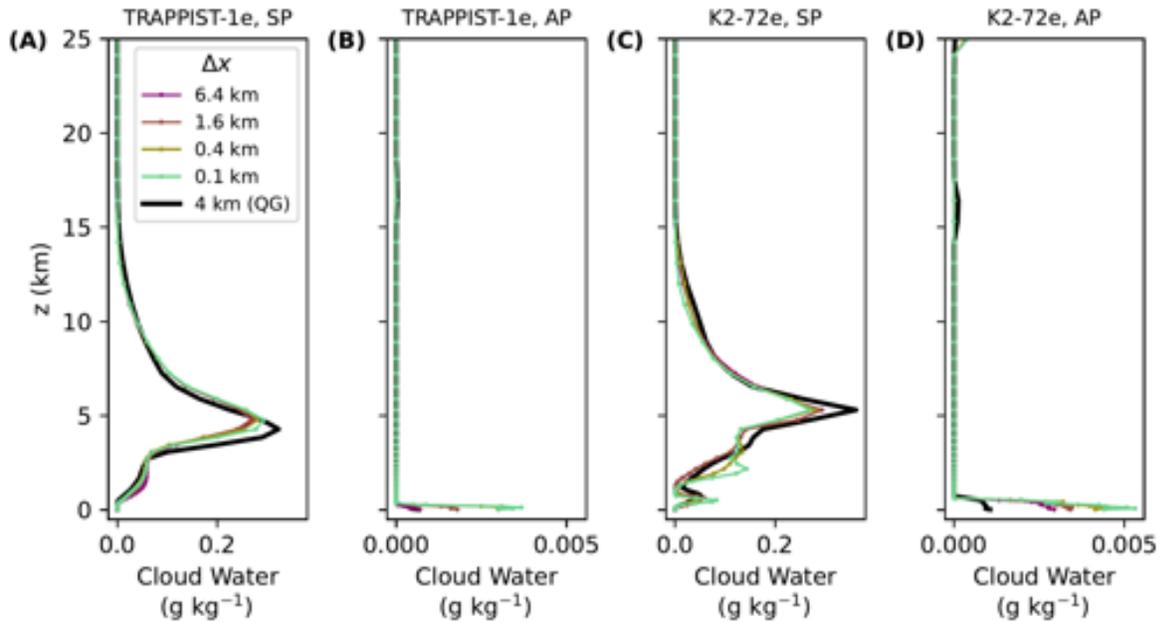

**Figure 3. Vertical profiles of cloud water content (ice + liquid) in the small-domain simulations** with different horizontal resolutions, 6.4, 1.6, 0.4, and 0.1 km. (A) and (B) are for TRAPPIST-1e, and (C) and (D) are for K2-72e. (A) & (C) are for the region around the substellar point (SP), and (B) and (D) are for the region around the antistellar point (AP). The black lines represent the quasi-global (QG) experiments with a grid spacing of 4 km. The SP profiles of cloud water content match well with the QG experiments, but the AP profiles with different resolutions have significant differences near the surface.

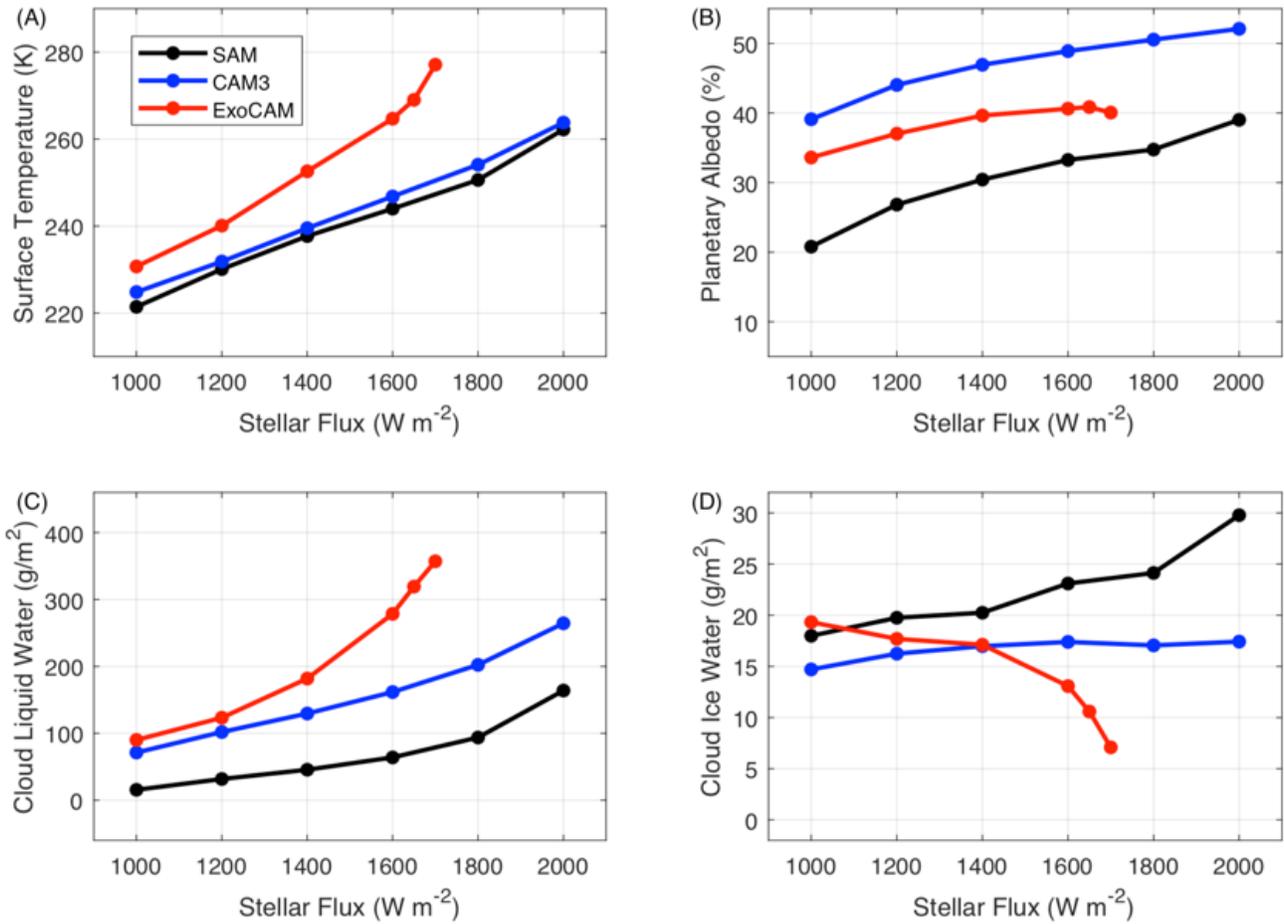

**Figure 4. Stabilizing cloud feedback simulated by the three models.** Time- and global-mean (A) surface temperature, (B) planetary albedo, (C) vertically-integrated cloud liquid water amount, and (D) vertically-integrated cloud ice water amount, as a function of the stellar flux at the substellar point. Black is for SAM, blue is for CAM3, and red is for ExoCAM. For SAM and CAM3, the stellar fluxes are 1000, 1200, 1400, 1600, 1800, and 2000 W m$^{-2}$, and the corresponding planetary rotation periods are 28.47, 24.83, 22.12, 20.01, 18.32, and 16.93 Earth days, respectively (following the Kepler's Third Law). For ExoCAM, the stellar fluxes are 1000, 1200, 1400, 1600, 1650, and 1700 W m$^{-2}$, and the corresponding planetary rotation periods are 28.47, 24.83, 22.12, 20.01, 19.56, and 19.12 Earth days, respectively. For SAM, the horizontal grid size is 20 km by 20 km. In these experiments, all the three models are coupled to a slab ocean but with zero ocean heat transport.

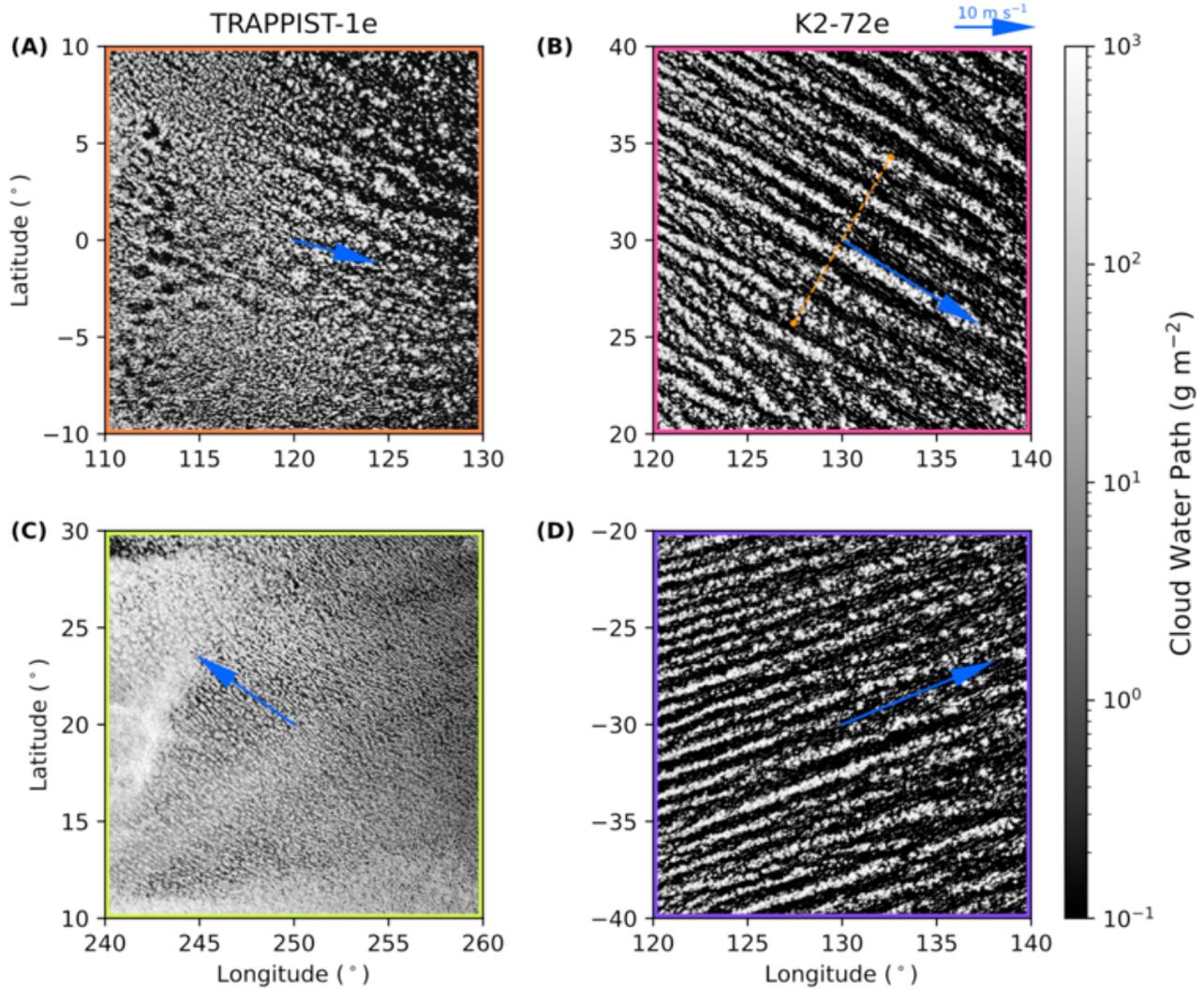

**Figure 5: Cloud streets in the SAM experiments with a resolution of 4 km.** The position and size of the selected region are indicated in Fig. 1 by the box with the same boundary colors. Left panels are for TRAPPIST-1e, and right panels are for K2-72e. The blue vector shows the direction and strength of the mean winds in the planetary boundary layer. The orange line in panel B is the region employed to analyze the formation of the cloud streets shown in Supplementary Fig. 17.

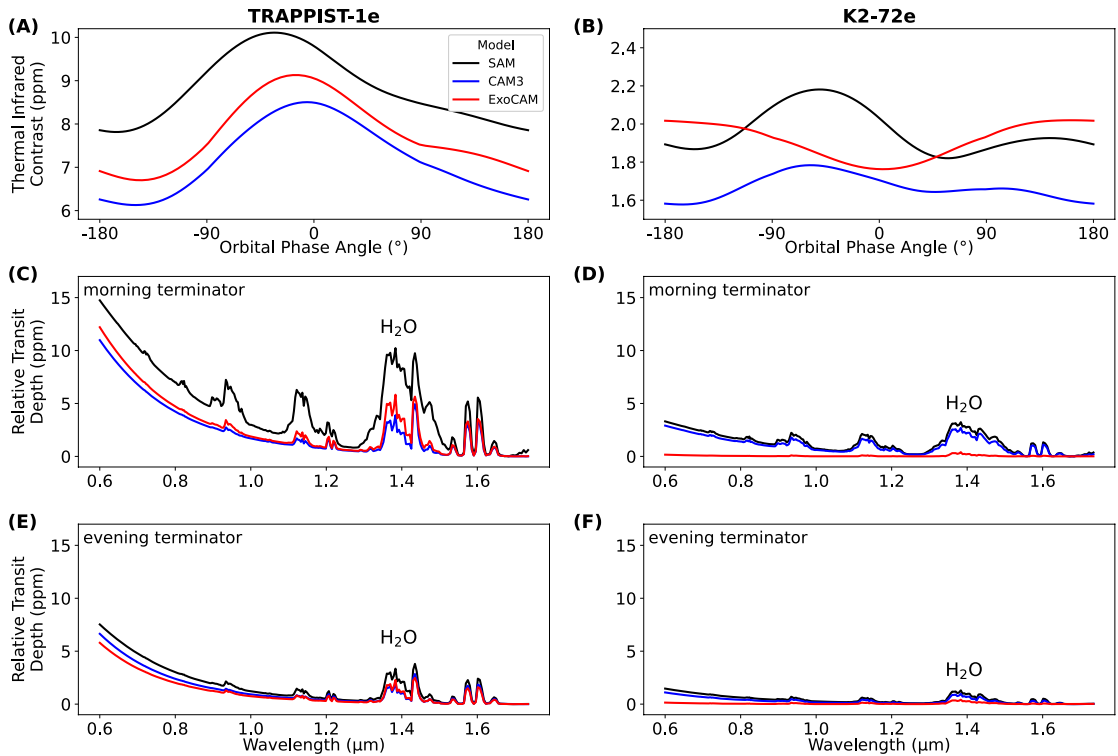

**Figure 6. The effects of cloud permitting on the observational characteristics of the planets.** Left is for TRAPPIST-1e, and right is for K2-72e. Panels A & B are for broadband thermal infrared emission phase curves; the $x$ axis is orbital phase angle; and the $y$ axis is the ratio of planetary thermal emission to stellar thermal emission (in 5-50 μm) in units of parts per million (ppm). One observer views the whole dayside at phase angle of 0° (the secondary eclipse) and sees the whole nightside at ±180° (transit). Observer inclination is assumed to be 90º. Note the different ranges of the $y$ axis between (A) and (B). Although K2-72e is warmer than TRAPPIST-1e, its thermal infrared contrast is lower, due to that the star K2-72 is ~7.6 times of TRAPPIST-1 in area. Panels C, D, E, & F are transmission spectra between 0.6 and 1.7 μm. Panels C & D are the morning terminator (90º longitude in Fig. 2), and panels E & F are the evening terminator (270º longitude). The minimum value for the relative transit depth has been subtracted for ease of display.

**This supporting information file includes:**
    Text 1: Previous simulations
    Text 2: Results of the SAM benchmark simulation for Earth
    Text 3: Different definitions for cloud fraction between different models
    Text 4: The sensitivity to vertical resolution and model top
    Text 5: Temperature limit for ice/liquid cloud formation
    Text 6: More details on the cloud streets
    Text 7: Atmospheric circulation, Cartesian geometry, and the limitations of this study
    Text 8: Convection and precipitation in extremely hot climates
    Text 9: Gibbs phenomenon in the CAM3 experiment of K2-72e
    References (27 in total)

    Supplementary Tables 1-4
    Supplementary Figures 1 to 21
    Legends for Supplementary Videos 1 to 3

**Other online supplementary materials for this manuscript include the following:**
    Supplementary Videos 1 to 3

**Text 1. Previous simulations.** Previous GCM inter-comparisons for tidally locked rocky planets have shown that there are large differences among GCMs, and the difference in global-mean surface temperature can be as large as 30 K when the models are run under the same boundary conditions (Yang et al. 2019; Fauchez et al. 2020; Sergeev et al. 2022a). Moreover, even in the simulations of present and future climates on Earth, clouds are the largest source of model uncertainty, since different GCMs employ different convection and cloud schemes (e.g., Cess et al. 1990; Zelinka et al. 2020). Given the lack of in-situ cloud observations for exoplanets, one of the best (although not perfect) solutions to this big problem is to use cloud-resolving models (CRMs) or cloud-permitting models (CPMs), which have fine spatial resolution and explicitly calculate convection and clouds without cumulus parameterization.

In the past five years, CRMs and CPMs have been employed to simulate convection/clouds on tidally locked planets but only in limited-area domains, such as 1000 by 1000 km (Zhang et al. 2017), an equatorial strip (Koll & Cronin 2017), 6000 by 6000 km (Sergeev et al. 2020), 250 by 250 km (Lefèvre et al. 2021), 72 by 72 km (Seeley & Wordsworth 2021), and a 2D idealized configuration along the equator (Song et al. 2022). While this small-domain approach is useful in examining the clouds in local regions, it is unable to investigate the clouds in a global view, to correctly represent the interactions between convection/clouds and large-scale dynamics, or to well examine the effect of clouds on planetary climate and habitability. Moreover, in the simulations of Zhang et al. (2017), Koll & Cronin (2017), Lefèvre et al. (2021), and Song et al. (2022), the important effect of the Coriolis force was not included. In order to overcome these shortcomings, here we carry out global-scale CPM simulations for tidally locked rocky planets orbiting around M dwarfs.

**Text 2. Results of the SAM benchmark simulation for Earth.** The results of the experiment testing the method of quasi-global cloud-permitting simulation is showed in Supplementary Fig. 6. Snapshots of cloud water path (liquid plus ice), surface precipitation, precipitable water (the sum of water vapor and liquid and ice clouds), and vertical velocity at the level of 5 km are shown in Supplementary Fig. 6(A-D). These figures show that the key characteristics of the Intertropical Convergence Zone (ITCZ) and mid-latitude baroclinic zone can be properly simulated. Supplementary Figs. 6(E-H) show the zonal-mean air temperature, specific humidity, zonal winds, and atmospheric mass streamfunction. These figures suggest that the model can suitably simulate the atmospheric stratification, water vapor field, zonal jets in mid-latitudes, tropical trade winds, and the Hadley cells, and the Ferrel cells.

Because the surface is ocean everywhere and there is neither seasonal cycle nor diurnal cycle, the model cannot well simulate the location of the ITCZ, the zonal asymmetry of the clouds such as stratocumulus clouds over eastern oceans, and stationary waves related to land-sea contrasts on Earth. For modern Earth, the mean location of the ITCZ is at the north of the equator, whereas in our aqua-planet simulation, the ITCZ is right at the equator. Supplementary Fig. 6(H) shows that there is another thin cell over the Hadley cell in each hemisphere; to our knowledge, the underlying reason is unclear yet but it may be related to the lack of the seasonal cycle in the experiment. Moreover, because two solid walls are set at the latitudes of 60°S and 60°N, some clouds are advected from lower latitudes and then trapped near the walls (Supplementary Fig. 6A & B). Meanwhile, an unrealistic overturning circulation near the wall is simulated in each hemisphere (Supplementary Fig. 6H). When the simulated region is extended to from 90°S to 90°N, the simulated atmospheric circulation is closer to that on Earth (Supplementary Fig. 7). The edges of the Hadley cells are at 30°S and 30°N, the Ferrel cells are right at the middle latitudes (~30°S-60°S and ~30°N-60°N), and the two polar cells are right at the high latitudes (~60°S-90°S and ~60°N-90°N).

**Text 3. Different definitions for cloud fraction between different models.** In SAM, cloud fraction is a diagnostic variable. It means that in the model source codes of SAM, cloud fraction is not required to be calculated. The cloud fraction shown in Supplementary Fig. 9 is diagnosed from the model output of cloud water content. At a given level, it is defined as 100% when the cloud water content is greater than 0.01 g kg$^{-1}$, otherwise it is 0% (Khairoutdinov & Randall 2003). The threshold of 0.01 g kg$^{-1}$ is in order to exclude very optically-thin clouds. For a given column from the surface to the top of the model, the value of cloud fraction is 100% when the vertically-integrated cloud water content is higher than 0.02 kg m$^{-2}$ (note it is different from the limit for a given level), otherwise it is 0%. These imply that for each time step and each grid the cloud fraction could be either 0% or 100%, not between. But, for long-term mean, it could be any value between 0% and 100%. For example, for a certain grid, if at one time step it is 100%, and at the next step it is 0%. Then, the average of these two steps is 50%.

Moreover, for cloud overlap between different levels, the overlap in SAM is 100% because the cloud (if it exists) occupies the entire grid. For each grid, there is no partial cloudy or clear sky, because the cloud is resolved in SAM. In contrast, there are partial cloudy and clear skies for each grid in GCMs such as CAM3 and ExoCAM. In GCMs, cloud fraction is also a diagnostic variable, but it is empirically parameterized based on relative humidity, convective mass flux, atmospheric stratification, and other variables (Collins et al. 2004). For a given level, it can be 0%, 100%, or any value between 0% and 100%. In GCMs, the overlap of clouds between different vertical levels could be minimum, maximum, random, or an arbitrary combination, depending on the parameterization scheme used.

**Text 4. The sensitivity to vertical resolution and model top.** In Lefèvre et al. (2021), the convective plumes reach 21 km when the sea surface temperature (SST) is ~320 K, 19 km at ~310 K, and ~17 km at ~300 K (see their Figure 2(a)). In our SAM simulations, the maximum SST is about 310 K and it occurs in only one experiment (see Supplementary Fig. 19(A)), and in other experiments the maximum SSTs are below 310 K or close to 300 K (Supplementary Fig. 8). In all our experiments, the sponge layer is between 18 and 27 km, so it does not significantly influence our results.

In order to further clarify this, we have done two additional experiments, within which the top of the model is extended and meanwhile the vertical resolution is increased. In one experiment, the top of the model is at 42 km, the number of the vertical levels is increased from 48 to 72, and the sponge layer is between 28 and 42 km. In the other experiment, the top of the model is at 65 km, the number of the vertical levels is 72, and the sponge layer is between 43 and 65 km. The result is showed in the following Supplementary Fig. 3. One could find that the results do not have essential changes although there are small detailed variations.

**Text 5. Temperature limits for ice/liquid cloud formation.** Different models use different temperature limits to distinguish between ice cloud and liquid cloud. In the three models (CAM3, ExoCAM, and SAM), the total cloud water is decomposed into ice and liquid clouds with assuming the ice cloud fraction is: $f_{ice} = (T - T_{max})/(T_{min} - T_{max})$, where $T$ is air temperature, $T_{max}$ is the maximum air temperature for ice cloud formation, and $T_{min}$ is the minimum air temperature for ice cloud formation. When $T$ is higher than $T_{max}$, all clouds are in liquid phase, and when $T$ is lower than $T_{min}$, all clouds are in ice phase.

Different models use different temperature limits. In CAM3 and ExoCAM, $T_{max}$ is equal to -10°C and $T_{min}$ is equal to -40°C. However, in SAM, $T_{max}$ is equal to 0°C and $T_{min}$ is equal to -20°C. In order to test these two parameters, we use CAM3 to do one experiment within which $T_{max}$ is changed from -10°C to 0°C and $T_{min}$ is changed from -40°C to -20°C. The results are shown in Supplementary Table 3 below, labelled as CAM3_TLC. We find that the vertically-integrated liquid cloud water path decreases but the vertically-integrated ice cloud water path increases, consistent with the change of the temperature limits. The planetary albedo changes from 47.0% to 41.0% and becomes closer to that of SAM. Other variables such as surface precipitation, cloud longwave radiative effect, and surface temperature also become closer to those in SAM. These results suggest that the temperature limits for liquid and ice clouds are two of the key parameters in explaining the differences between SAM and CAM3/ExoCAM; of course, other parameters should also influence the simulation results.

**Text 6. More details on the cloud streets:** The cloud streets are parallel bands of low-level cumulus clouds in convective boundary layer, and the cloud bands are oriented nearly parallel to the mean winds. On Earth, all clouds are generally classified as ten mutually exclusive cloud genera, and the cloud streets belong to the genus of stratocumulus (Houze 2014). One necessary condition for the formation of cloud streets on tidally locked planets is the large day-night surface temperature contrast. Because the contrast decreases with increasing stellar flux (Haqq-Misra et al. 2018), there should have few or no cloud streets for planets close to the inner edge of the habitable zone. For planets close to the outer edge of the habitable zone and having dense $CO_2$ or other greenhouse gases, the day-night surface temperature is also smaller, but it is still larger than such as ~20 K under a 20 bar $CO_2$ atmosphere (see Fig. 1 in Wordsworth et al. (2011)), allowing the possible formation of cloud streets.

In our experiments, the relatively cloudless regions between the cloud streets (see Supplementary Fig. 17 below) can be viewed as radiator fins. Radiator fins represent clear-sky dry columns those have relatively large outgoing longwave radiation to space. This can be compared with either cloud-sky columns or clear-sky wet columns. Both cloud-sky and clear-sky wet columns have lower outgoing longwave radiation than that of adjacent, clear-sky dry columns. Pierrehumbert (1995) called the dry columns of the subtropics as "radiator fins". Please see Fig. 7 in Pierrehumbert (1995) for a schematic representation for the radiator fin.

On Earth, the lifetime of cloud streets is always in the range between hours and several days, mainly determined by the duration of cold air outbreaks (e.g., Brummer & Pohlmann, 2000; Gryschka &

Raasch 2005). On tidally locked planets, the cloud streets persist in the entire time of the simulations, so that the cloud streets are not hard to see even in the time-mean field of cloud water amount shown in Fig. 2A & B. This is due to the fact that the favorable conditions for the formation of the cloud streets, such as the strong night-to-day cold advection and the temperature inversion, never stop on tidally locked planets. Moreover, due to the large distance of the cold advection from the nightside to the dayside, the cloud bands along the mean winds is long in length, $\sim 10^3$ to $10^4$ km (see Fig. 1 in the main text).

Besides the cloud streets, one could also see gravity-wave clouds (Supplementary Videos 1-3). The gravity-wave cloud bands are nearly perpendicular to the direction of large-scale mean winds, and the gravity waves propagate downstream along the mean winds.

Moreover, the atmospheric composition is $N_2$-dominated in this study, which is heavier than $H_2O$ in molecular weight and thereby promotes convection (such as Li & Ingersoll 2015). If the atmosphere is $H_2$- or He-dominated, a moist parcel would be heavier than a dry parcel under the same pressure and temperature, and therefore moist convection would likely be inhibited and both cloud streets and deep convective clouds would be much harder to form.

**Text 7. Atmospheric circulation, Cartesian geometry, and the limitations of this study.** SAM is able to basically simulate the large-scale atmospheric circulation. Strong upwelling occupies the substellar region and the rest of the planet is mainly dominated by weak downwelling. The upwelling and downwelling are linked by strong convergence in the planetary boundary layer towards the substellar region and strong divergence in the free troposphere towards the nightside and the high latitudes of the dayside (Supplementary Figs. 11 & 12). This circulation is called as a global-scale overturning or 'Walker' circulation (e.g., Showman et al. 2013; Hammond & Lewis 2021). In the horizontal plane of the free troposphere, the atmosphere is characterized by an equatorial superrotation flowing from west to east (Supplementary Fig. 12). The superrotation is maintained by equatorward momentum transports by coupled Rossby-Kelvin waves, ultimately driven by the uneven stellar radiation distribution between the permanent dayside and nightside (e.g., Showman & Polvani 2011). Note that there are dramatic differences in the spatial pattern of the convergence/divergence between SAM and CAM3/ExoCAM (Supplementary Fig. 11), so that the distributions of cloud water on the dayside are quite different (see Fig. 2 in the main text).

Supplementary Fig. 8 shows the surface temperatures simulated by SAM and the two general circulation models CAM3 and ExoCAM. For TRAPPIST-1e, the spatial pattern of the surface temperatures is similar between the three models. In global mean, the surface temperatures are similar, 243, 240, and 243 K in SAM, CAM3, and ExoCAM, respectively. For K2-72e, the spatial pattern of the surface temperatures has a significant difference: In SAM, the west side of the substellar point is significantly warmer than the east side of the substellar point (Supplementary Fig. 8B), but the symmetry around the substellar point is almost perfect in CAM3 and ExoCAM. The strong zonal asymmetry in SAM can also be viewed from the spatial distributions of vertical velocity and horizontal winds of K2-72e as shown in Supplementary Figs. 11-12. This is likely due to the interaction among the resolved clouds, radiative transfer, and equatorial superrotation. The deep convective clouds are transported to the east side of the substellar point by the equatorial superrotation, so the clouds and planetary albedo exhibit strong zonal tilts towards the east (see Fig. 2B in the main text). Therefore, much more shortwave radiation reaches the west surface of the substellar point than that on the east side (Supplementary Fig. 13). As a result, the west side is warmer than the east side. This zonal asymmetry is much weaker in CAM3 and ExoCAM. However, the global-mean surface temperatures have relatively small differences, 263, 253, and 268 K in SAM, CAM3, and ExoCAM, respectively.

The dayside and the low- and middle-latitudes of the nightside in SAM are slightly warmer than CAM3, but the high latitudes in SAM are somewhat cooler. For night-side mean, the surface temperature of TRAPPIST-1e is 219.4 and 215.9 K in SAM and CAM3, respectively, and the surface temperature of K2-72e is 241.4 and 227.5 K in SAM and CAM3, respectively. One of the reasons for the differences is related to water vapor concentration on the night side. The mean night-side water vapor concentrations of TRAPPIST-1e are respectively 2.6 and 1.9 kg m$^{-2}$ in SAM and CAM3. For K2-72e, these two values are 8.3 and 4.1 kg m$^{-2}$, respectively (Supplementary Fig. 14). The differences in water vapor concentration act to make the night-side surface of SAM be relatively warmer than CAM3. A part of the water vapor is from local evaporation, and the rest is from horizontal transport from the dayside (Supplementary Table 4).

In SAM, the simulated equatorial superrotation is stronger than those obtained in CAM3 and ExoCAM, especially for K2-72e (Supplementary Fig. 12). Several reasons can cause this. (1) Latent heat release (can be viewed from the surface precipitation shown in Supplementary Fig. 10J) in SAM is larger than those in CAM/ExoCAM. The global-mean surface precipitation rates on K2-72e are 2.4, 1.1, and 0.9 mm day$^{-1}$ in SAM, CAM3, and ExoCAM, respectively. More latent heat release (or more accurately larger gradients in the spatial pattern of the latent heat release) can drive larger perturbation in the geopotential field of the atmosphere and subsequently induce stronger waves and equatorial superrotation. (2) The Rossby deformation radius of the atmosphere of TRAPPIST-1e is at the edge between fast and Rhines rotation regimes, so a change in convection scheme, initial condition, boundary-layer friction or other factor can push the atmospheric circulation from one regime to the other regime (e.g., Sergeev et al. 2020, 2022a, 2022b; Haqq-Mirsa et al. 2018; Carone et al. 2016; Edson et al. 2011). Convection scheme also has strong effects on slowly rotating planets such as Proxima Centauri b (Sergeev et al. 2020). (3) The models employ different damping schemes near the top of the model (numerical dissipation or sponge), and this can influence the circulation in the upper atmosphere (Turbet et al. 2022). (4) The models use different dynamical cores (Eulerian in CAM3, finite volume in ExoCAM, and finite difference in SAM), which can also influence the simulated atmospheric circulation (e.g., Lee & Richardson 2010). (5) SAM in these simulations employs a global Cartesian geometry rather than the realistic global spherical geometry as that used in CAM3 and ExoCAM. This can influence the atmospheric circulation especially in the high latitudes.

From Supplementary Fig. 12, one could find that in the free troposphere there is one cyclone in the region between 0° and 90° longitudes of each hemisphere for both CAM3 and ExoCAM. This is consistent with previous theoretical derivations, 2D shallow water model studies, and 3D GCM simulations (e.g., Showman & Polvani 2011; Tsai et al. 2014; Hammond & Pierrehumbert 2018; Wang & Yang 2021). SAM tends to produce the cyclones but the spatial pattern is not obvious. This is likely due to that a global-scale Cartesian geometry rather than a global spherical geometry is used in the simulations (see Methods), so the simulated high-latitude atmospheric circulation has its own flaws in SAM.

How strong does the Cartesian geometry influence the simulation results in this study? We answer this question based on our best understanding. (1) The Cartesian geometry has two weaknesses: the grid area in high latitudes is larger than that of the spherical geometry, and the momentum equations do not include the metric terms (or curvature effects). The former can influence the atmospheric circulation in high latitudes. The magnitude of the geometric terms is always small as long as the winds are smaller than the order of 100-1000 m s$^{-1}$ (Chapter 2 in Holton & Hakim (2012)). In our simulations of both aqua-Earth and tidally locked rocky planets, the wind speed is lower than the limit. (2) The planetary rotation is central for many dynamical phenomena such as Hadley cells, Ferrel cells, baroclinic instability, Rossby waves, Kelvin waves, and wave—mean-flow interactions, but the sphericity of the planet is not always important (Vallis 2019). Since the pattern of the Coriolis parameter in the quasi-global simulations is the same as the real pattern, the key behaviors of the atmospheric circulation should be correctly simulated. (3) Local convective-scale motion can be well captured by the Cartesian geometry, if large-scale forcing is correctly given. This is because the radii of Earth and other terrestrial planets are much larger than the characteristic length of convection. This is also the reason why the Cartesian geometry is so widely used in small-domain cloud-solving simulations. (4) Conservation properties are not sacrificed in the Cartesian geometry. Key thermodynamical and dynamical properties, including momentum, mass, and moist static energy, are conserved. This means that the basic laws of fluid dynamics are well represented in our quasi-global simulations. (5) Our simulations are able to capture the key features of the large-scale circulation on tidally-locked planets, including superrotation and the global 'Walker circulation' (see Supplementary Figs. 11 & 12). These features are originated mainly from planetary rotation and the zonal variation of the stellar forcing. The distribution of the incident radiation in our simulations matches that in the spherical geometry, both featuring a strong heating near the substellar point and zero incident radiation at the night side. (6) The Cartesian geometry cannot well represent the circulation at the high latitudes. So, we make clouds and the circulation at low and middle latitudes be the focus of this study and admit that the circulation at high latitudes in our simulations might not be correct. (7) In Hammond and Pierrehumbert (2018), they compared the effect of changing the geometry from a beta-plane Cartesian to a sphere in a linear shallow water model for tidally locked planets, and they found that "the beta-plane approximation produces much the same results as the spherical geometry", although small detailed differences exist (see the Fig. 6 versus Fig. 10 in their paper).

Overall, the use of the Cartesian geometry rather than the spherical geometry is a limit of this study, but in theory this is not a big or fatal problem and will not influence the main conclusions. Meanwhile, our group is trying to re-do the simulations using other two models WRF and MPAS. These two models use spherical geometry. We will know more about this in the near future.

Another limitation of this study is that there is no cross-pole flow in our SAM experiments, due to the use of rigid wall boundary condition at the two poles. Previous studies have showed that there are significant cross-pole flows on tidally locked planets, as showed in Figures 10-12 in Joshi et al. (1997), Figures 6-8 in Haqq-Misra & Kopparapu (2015), and Figure 7 of Kopparapu et al. (2016). How large does this shortcoming of SAM influence the results? In order to answer this question, we calculate the day-to-night mass transports. The total day-to-night transport in ExoCAM is contributed by two parts, cross-pole flows and cross-terminator flows. In SAM, all the day-to-night transport is from cross-terminator flows. In supplementary Table 4, we compare the net day-to-night transports of water vapor and clouds between SAM and ExoCAM. For the water vapor transport, SAM is relatively smaller than ExoCAM in the simulation of TRAPPIST-1e, 10.6 versus 19.1 kg m$^{-1}$ s$^{-1}$, but larger in the experiment of K2-72e, 33.8 versus 30.6 kg m$^{-1}$ s$^{-1}$. For the former, it is at least partially related to the lack of cross-pole flow in the SAM experiment. For the latter, although there is no cross-pole flow in SAM, the zonal flows in the SAM experiment of K2-72e are relatively stronger than that in ExoCAM (see supplementary Figs. 12B & 12F), which can make the day-to-night water vapor transport in SAM be relatively greater. For the cloud water transport from dayside to nightside, SAM is smaller than ExoCAM in the simulation of TRAPPIST-1e, 0.1 versus 0.6 kg m$^{-1}$ s$^{-1}$, but larger in the experiment of K2-72e, 0.8 versus 0.3 kg m$^{-1}$ s$^{-1}$. These trends are the same as those for water vapor, and the reasons should be similar because these two traces (clouds and water vapor) are transported by the same winds. Moreover, the dayside cloud water amount of TRAPPIST-1e in the SAM experiment is less than that in ExoCAM (Fig. 2 in the main text); this can also reduce the magnitude of the day-to-night cloud transport in SAM. Therefore, the lack of cross-pole winds in SAM is one of the reasons for the relatively less night-side cloud cover (Supplementary Fig. 9) in the simulation for TRAPPIST-1e, but not for K2-72e. The key reason is likely the much less cloud formation on the night side in SAM; this can be viewed from the supplementary videos online.

**Text 8. Convection and precipitation in extremely hot climates.** Using three different cloud-resolving models (DAM, CM1, and SAM), Seeley & Wordsworth (2021) investigated the convection behaviour under extremely hot climates above 320 K. Three domain sizes were employed, 72 km × 72 km, 144 km × 144 km, and 512 km × 512 km. They found that in the extremely hot climates, the system enters one regime called "episodic deluges": periodic, short, and strong precipitations are separated by relatively long dry spells. The strength of precipitation can reach tens of mm day$^{-1}$ or even over 100 mm day$^{-1}$. The period of the oscillations is about 1 to 4 Earth days.

In our quasi-global cloud-permitting simulations for tidally locked planets, we find a similar phenomenon (Supplementary Fig. 19). Supplementary Fig. 19(b) shows that the strength of precipitation can reach 100-200 mm day$^{-1}$ in a short time, and during the dry spells the precipitation is close to zero. But, the oscillations are much less regular than those found in the small-domain simulations of Seeley &

Wordsworth (2021); this makes it hard to verify what exact period it is in our experiments. Moreover, it is important to point out that this oscillation behavior occurs only in a small area in our simulations, neither the whole dayside nor a wide region around the substellar point.

In Seeley & Wordsworth (2021), they suggested that positive net lower-tropospheric radiative heating (LTRH) is the main mechanism that stabilizes the lower troposphere and decouples the surface and the upper troposphere during the dry spells, and the trigger of intense convection and precipitation is related to evaporative cooling at the base of elevated convection. LTRH causes an inhibition layer in the lower troposphere. When the evaporative cooling erodes the inhibition layer, deep convection starting from the near surface is allowed and strong rainfall occurs.

In our simulations, the mechanism is likely different from that showed in Seeley & Wordsworth (2021). As shown in Supplementary Fig. 20(C), the radiative (shortwave plus longwave) heating rate is positive in the whole troposphere at the selected region. This is due to the concentrated stellar radiation on the dayside. The star temperature is 3300 K, so more stellar energy is in the range of near-infrared wavelengths, within which water vapor and clouds can absorb more stellar energy. Note that in Seeley & Wordsworth (2021), the radiative heating rate is positive only in the near-surface layers (see their Figure 2b & c). Moreover, during the deluges in our experiment, the convection top reaches only ~8 km, rather than the whole troposphere (Supplementary Fig. 20). Supplementary Fig. 20(D) shows that there are two separated layers in the atmosphere: one is between the near surface and the level of ~8 km, and the other one is between 10 and 20 km. These two convection layers seem have no direct connection during all the time of the simulation. This is quite different from that found in Seeley & Wordsworth (2021). In their simulations, the convection during the deluge phase occupies the whole troposphere from the near surface to the tropopause.

All the above differences suggest that the underlying mechanisms are likely quite different. The apparent reason is that large-scale circulation is included in our simulations but not in Seeley & Wordsworth (2021). Further overthought analyses are required and a separated article is required to clearly address them.

**Text 9. Gibbs phenomenon in the CAM3 experiment of K2-72e.** The oscillation of cloud fraction in the CAM3 simulation (Supplementary Fig. 9(D)) can be explained by Gibbs phenomenon, which refers to the overshoot/undershoot of a partial sum expansion of a function near a discontinuity as compared to the original function (Navarra et al. 1994; Raeen 2008). In the CAM3 simulations, we use the spectral Eulerian dynamical core and the T31 truncation (Collins et al. 2004). The spectral method has the problem of producing artificial oscillation at discontinuities, especially when the horizontal resolution is not high.

Here, we demonstrate the Gibbs phenomenon by applying the T31 truncation to a simple 2D field that is 1 on the dayside and -1 on the nightside. The field truncated at T31 is shown in Supplementary Fig. 21(C), and ripple-like oscillations can be seen in the truncated field. The wavelength of the Gibbs oscillation is the circumference of the planet divided by 32. This value is approximately the shortest resolved zonal wave at the equator (Laprise 1992). The truncated pattern resembles the oscillations in the spatial pattern of low-level cloud fraction (Supplementary Fig. 21(A)).

We have also added one CAM3 experiment using the spectral Eulerian dynamical core under a T42 truncation. The result of low-level cloud fraction is showed in Supplementary Fig. 21(B). A truncated pattern of T42 (Supplementary Fig. 21(D)) also resembles the oscillations in the spatial pattern of the simulated low-level cloud fraction.

**Table 1.** Planetary properties of two rocky planets (refs. 39,40,73,74).

| Planet | TRAPPIST-1e | K2-72e |
|---|---|---|
| Stellar flux (W m$^{-2}$) | 900 | 1510 |
| Stellar temperature (K) | 2516 | 3360 |
| Planetary radius ($R_\oplus$) | 0.91 | 1.29 |
| Surface gravity ($g_\oplus$) | 0.93 | 1.29 |
| Orbital period (Earth days) | 6.1 | 24.2 |
| Surface pressure (bar) | 0.93 | 1.29 |

**Table 2.** Experimental configuration of cloud-permitting simulations

| Model | SAM |
| --- | --- |
| East-west domain extent | 38,400 km for TRAPPIST-1e, and 51,200 km for K2-72e |
| North-south domain extent | 19,200 km for TRAPPIST-1e, and 25,600 km for K2-72e |
| Horizontal grid spacing | 4 km by 4 km or 2 km by 2 km |
| Coriolis parameter | $f = 2\Omega\sin\varphi$, where $\Omega$ is rotation rate and $\varphi$ is latitude |
| Vertical grid | 48 layers extending to 27 km |
| Time step | 10 or 20 s |
| Lateral boundary condition | North & South: free-slip wall (there is no cross-pole transport in the model); East & West: periodic |
| Top boundary condition | Rigid lid at the top of the model and damping layers above ~18 km |
| Control experiments | Surface temperatures are specified from the equilibrium results of slab ocean experiments using SAM under a horizontal grid spacing of 40 km by 40 km. The simulation length is 210 Earth days (under 40 km) + 50 Earth days (under 4 km) or + 30 Earth days (under 2 km, Supplementary Fig. 1). |
| Stabilizing cloud feedback | Six different stellar fluxes are examined. The stellar temperature is 3300 K, and Earth's gravity and radius are used. Horizontal grid spacing is 20 km by 20 km, and the domain size is 20,000 km by 40,000 km. The surface is coupled to a slab ocean with a depth of 1.0 m. The simulation length is 350 Earth days (Supplementary Fig. 5). |

**Table 3**. Summary of main characteristics in the simulations of TRAPPIST-1e and K2-72e by the three models of SAM, CAM3, and ExoCAM. CAM3_TLC is a sensitivity test within which the maximum air temperature for ice clouds is changed from -10°C to 0°C, and the minimum air temperature for ice clouds is changed from -40°C to -20°C. The upper part for global mean, and the lower part is for night-side mean.

| Planet | TRAPPIST-1e | | | K2-72e | | | |
|---|---|---|---|---|---|---|---|
| Model | SAM | CAM3 | ExoCAM | SAM | CAM3 | CAM3_TLC | ExoCAM |
| *Global-mean values* | | | | | | | |
| Surface temperature (K) | 243.4 | 239.7 | 242.6 | 262.5 | 252.7 | 261.0 | 267.6 |
| Planetary albedo (%) | 18.1 | 29.1 | 22.1 | 37.4 | 47.0 | 41.0 | 39.0 |
| Water vapor (kg m$^{-2}$) | 6.0 | 4.5 | 7.7 | 15.6 | 11.6 | 19.7 | 29.6 |
| Surface precipitation (mm day$^{-1}$) | 1.8 | 1.4 | 1.2 | 2.4 | 1.1 | 1.6 | 0.9 |
| Total cloud fraction (%) | 27 | 95 | 90 | 41 | 82 | 87 | 97 |
| Total cloud water path (g m$^{-2}$) | 53.4 | 89.6 | 130.3 | 159.2 | 164.6 | 142.7 | 221.0 |
| Cloud liquid water path (g m$^{-2}$) | 26.1 | 65.5 | 107.8 | 136.1 | 152.5 | 126.4 | 210.8 |
| Cloud ice water path (g m$^{-2}$) | 27.3 | 24.1 | 22.5 | 23.1 | 12.1 | 16.3 | 10.2 |
| Longwave cloud effect (W m$^{-2}$) | 11.4 | 16.4 | 13.4 | 13.4 | 18.6 | 18.3 | 20.6 |
| *Night-side mean values* | | | | | | | |
| Surface temperature (K) | 219.4 | 215.9 | 219.1 | 241.4 | 227.5 | 237.3 | 252.3 |
| Planetary albedo (%) | 0 | 0 | 0 | 0 | 0 | 0 | 0 |
| Water vapor (kg m$^{-2}$) | 2.6 | 1.9 | 3.3 | 8.3 | 4.1 | 8.0 | 20.1 |
| Surface precipitation (mm day$^{-1}$) | 0.19 | 0.14 | 0.29 | 0.28 | 0.05 | 0.09 | 0.28 |
| Total cloud fraction (%) | 20 | 98 | 98 | 25 | 70 | 80 | 98 |
| Total cloud water path (g m$^{-2}$) | 9.0 | 23.9 | 58.9 | 55.9 | 5.7 | 6.6 | 94.3 |
| Cloud liquid water path (g m$^{-2}$) | 0.4 | 11.3 | 42.8 | 49.7 | 3.6 | 3.7 | 90.4 |
| Cloud ice water path (g m$^{-2}$) | 8.6 | 12.6 | 16.1 | 6.2 | 2.1 | 2.9 | 3.9 |
| Longwave cloud effect (W m$^{-2}$)[a] | -4.2 | -1.1 | -4.5 | -3.8 | -2.8 | -1.9 | -5.2 |

a. Note that the night-side cloud longwave radiative effect is negative rather than positive; this is due to that the low-level clouds locate near the top of the nightside temperature inversion, so that these clouds increase thermal emission to space and cool the atmosphere and surface (see also Fig. 3(b) in Yang et al. (2019)).

**Table 4**. Net vertically-integrated day-to-night mass transports in the simulations of TRAPPIST-1e and K2-72e by SAM and ExoCAM. The cloud water transport includes both liquid and ice clouds. These transports are calculated along the great circle consisting of the west and east terminators, and the mean value along the great circle is listed here. There is cross-pole transport in ExoCAM but no in SAM. Note the large differences between the two models.

| Planet | TRAPPIST-1e | | K2-72e | |
|---|---|---|---|---|
| Model | SAM | ExoCAM | SAM | ExoCAM |
| water vapor transport (kg m$^{-1}$ s$^{-1}$) | 10.6 | 19.1 | 33.8 | 30.6 |
| cloud water transport (kg m$^{-1}$ s$^{-1}$) | 0.1 | 0.6 | 0.8 | 0.3 |

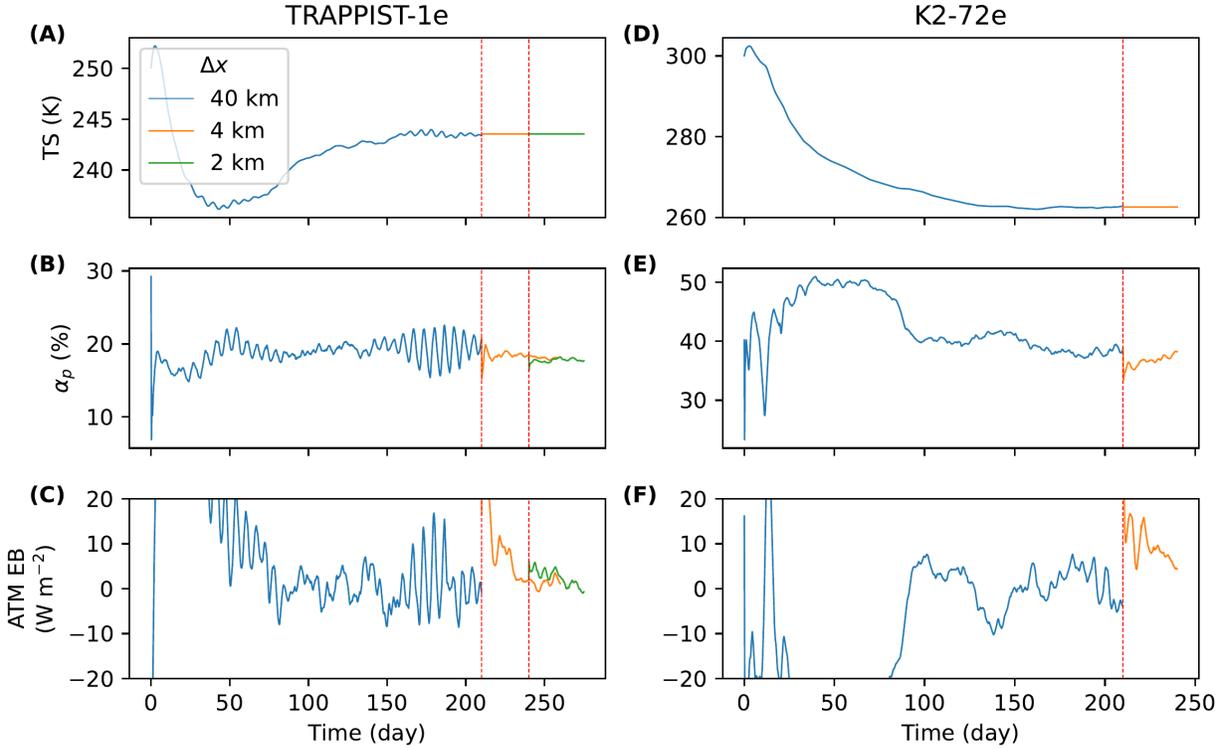

**Figure 1. Time series of global-mean surface temperature (A & D), planetary albedo (B & E), and atmospheric energy balance (ATM EB, (C & F)) in the SAM experiments.** Left panels are for TRAPPIST-1e, and right panels are for K2-72e. As described in Methods, the first 210 days (blue line) are under a horizontal resolution of 40 km by 40 km and coupled to a slab ocean, the yellow line is for a resolution of 4 km by 4 km but with fixed sea surface temperatures (SSTs), and the green line is for a resolution of 2 km by 2 km also with fixed SSTs. ATM EB is defined as the net flux between energy input and energy output to the atmosphere; energy input includes shortwave absorption by the atmosphere and clouds, surface sensible heat, surface latent heat, and longwave radiation from the surface; and energy output includes longwave radiation to space and to the surface. Note the large oscillations of planetary albedo and ATM EB during days ~30-70 and 160-200 in the experiment of TRAPPIST-1e; it may be due to the reflection of gravity waves near the bottom of the model's sponge layer; the reflected gravity waves strongly disturb the clouds in the troposphere and affect the planetary albedo; this is unrealistic because the gravity waves could propagate to higher altitudes if the model top were higher in altitude or the model were more realistic in the top boundary condition. Due to computation source limitation, the 4 km and 2 km experiments use fixed SSTs; the experiments have reached quasi-equilibrium but not perfect equilibrium; further simulations under a small domain (Fig. 3 in the main text) suggest that our main conclusions are robust.

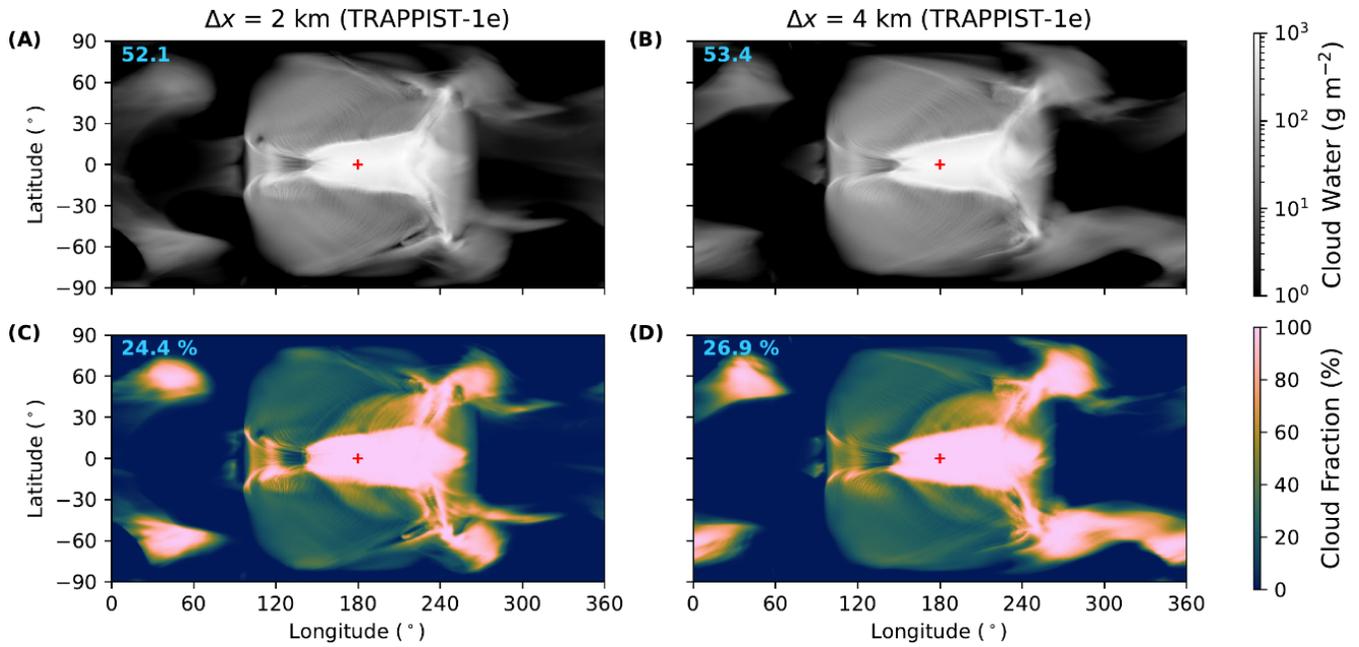

**Figure 2. A comparison of cloud patterns in the TRAPPIST-1e simulations between grid spacing of 2 km (left panels) and of 4 km (right panels).** Panel A and B show the time-mean cloud water path (g m$^{-2}$) in the last 10 days. Panel C and D show the cloud fraction (%) averaged during the same the time period. The number at the upper left corner in each panel is the global-mean value. The red cross marks the substellar point.

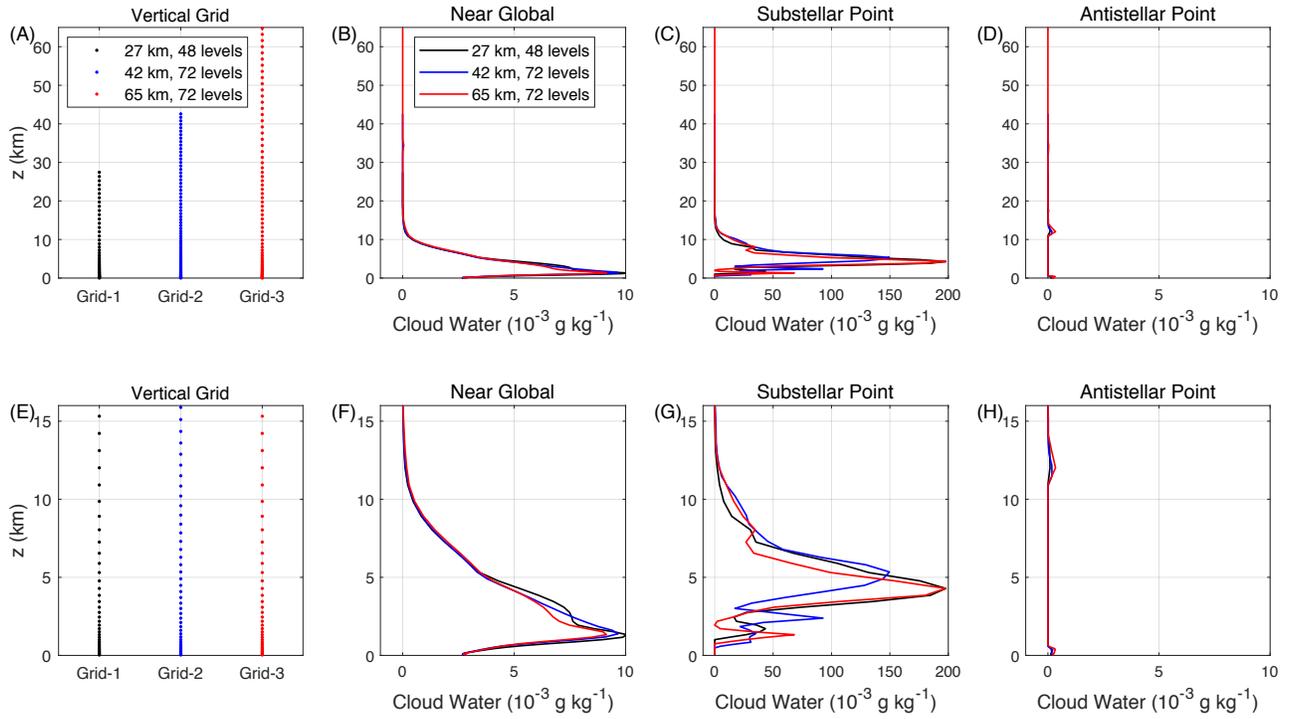

**Figure 3. The results of changing vertical resolution and extending the top of the model using SAM.** Upper panels: (A) vertical grids, (B) global-mean cloud water path, (C) cloud water path over the substellar point, and (D) cloud water path over the antistellar point. Bottom panels: same as the upper panels, but only the below 16 km is showed. In these simulations, the stellar flux at the substellar point is 1400 W m$^{-2}$, the rotation period (= orbital period) is 22.12 Earth days, and the horizontal resolution is 11 km by 14 km over a quasi-global domain.

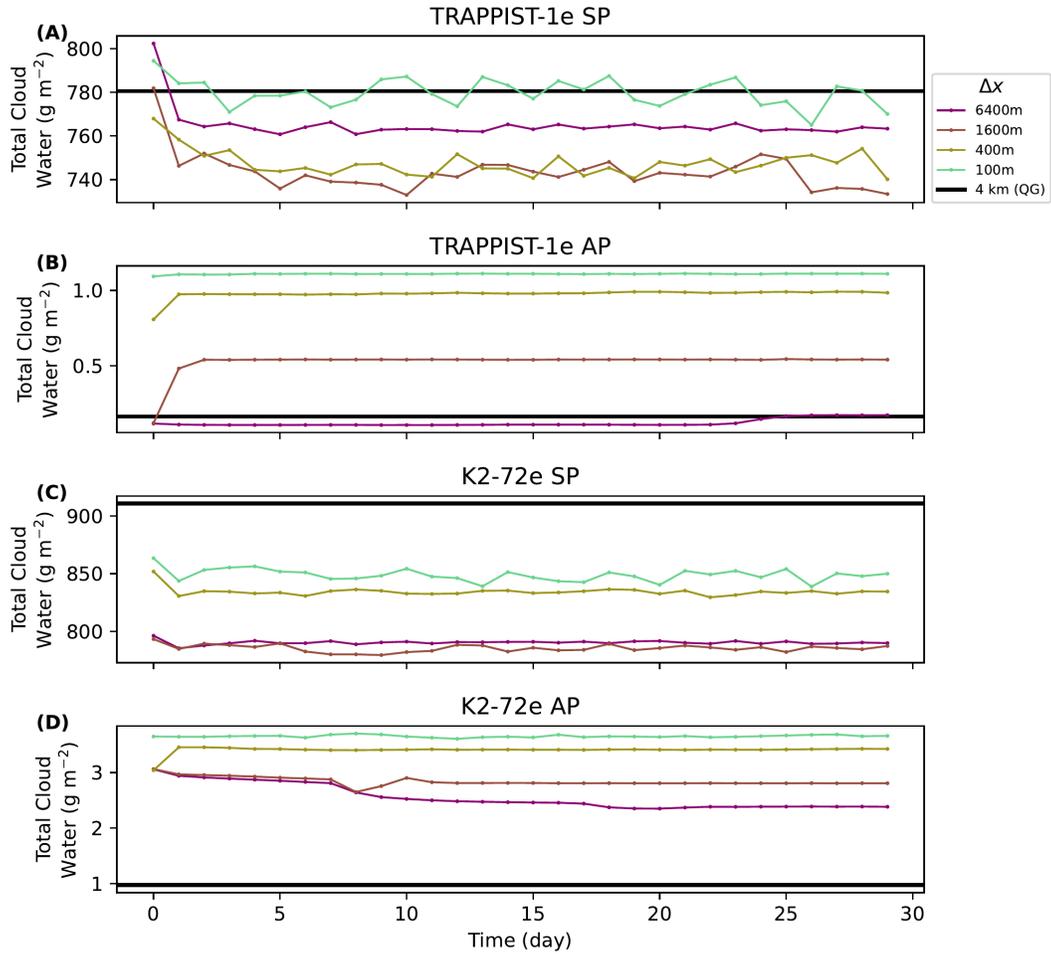

**Figure 4**. **Time series of vertically-integrated, domain-mean total (liquid + ice) cloud water path at the substellar point (SP) and the antistellar point (AP) for TRAPPIST-1e (A & B) and K2-72e (C & D) in the small-domain experiments**. The horizontal resolution is 6.4, 1.6, 0.4, or 0.1 km. In all these experiments, the grid numbers are 32 by 32, so the domain sizes are different between the experiments. The results of the quasi-global (QG) simulations are also showed. Between different resolutions, the total cloud water path changes relatively small at the SP but relatively large at the AP.

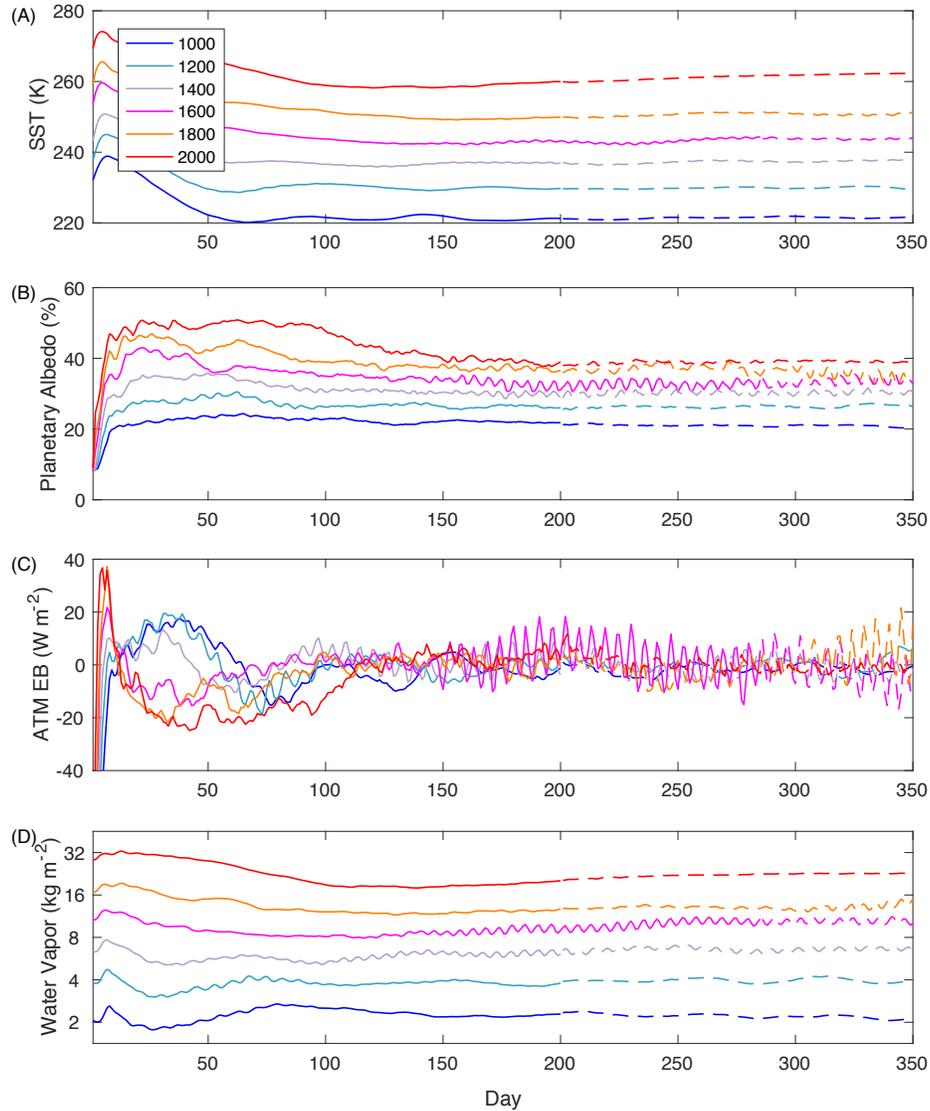

**Figure 5**. **Time series of global-mean surface temperature (A), planetary albedo (B), atmospheric energy balance (ATM EB, (C)), and vertically-integrated water vapor (D) in the stabilizing cloud feedback experiments using SAM**. Six different stellar fluxes at the substellar point are examined, 1000, 1200, 1400, 1600, 1800, and 2000 W m$^{-2}$, and the corresponding planetary rotation periods are 28.47, 24.83, 22.12, 20.01, 18.32, and 16.93 Earth days, respectively (following the Kepler's laws). The stellar temperature is 3300 K, and Earth's gravity and radius are used in these experiments. The horizontal resolution in these experiments is 40 km by 40 km for the periods of solid lines but 20 km by 20 km for the periods of dashed lines. The model is coupled to a slab ocean with a depth of 1.0 m.

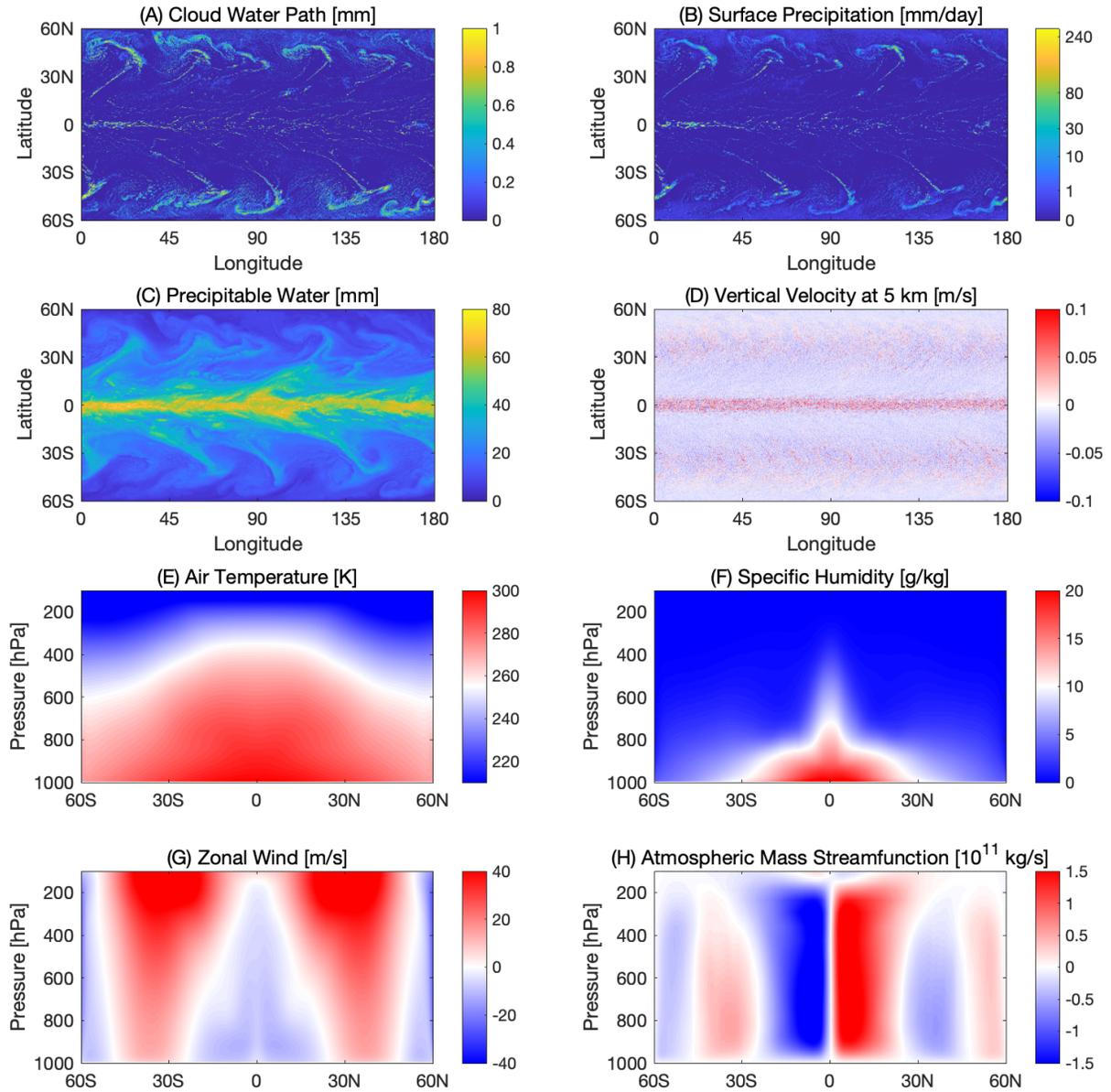

**Figure 6. Climatic characteristics in the SAM benchmark simulation.** Panels A to D: snapshots of cloud water path (liquid plus ice), surface precipitation, precipitable water (the sum of water vapor and liquid and ice clouds), and vertical velocity at the level of 5 km, respectively. Panels E to H: the 30-days mean zonal-mean air temperature, specific humidity, zonal winds, and atmospheric mass streamfunction. In the experiment, solar constant is 1360 W m$^{-2}$, $CO_2$ concentration is 369 ppmv, rotation period is 1 Earth day, orbital period is 365 Earth days, planetary obliquity is 0°, and orbital eccentricity is 0°, and the surface is ocean everywhere.

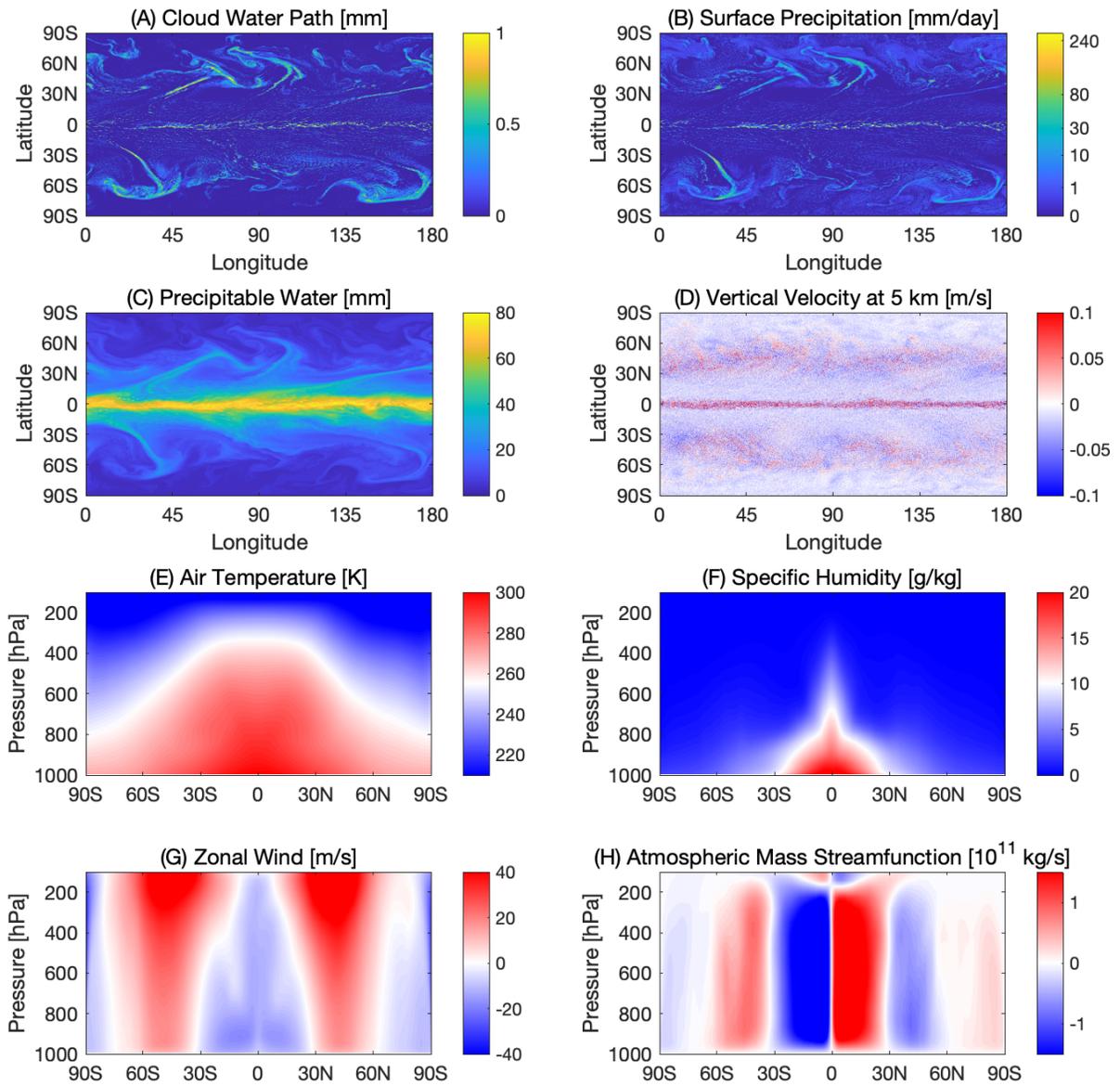

**Figure 7.** Same as the supplementary Figure 6 above, but for the experiment within which the southern and northern boundaries are moved from 60°S and 60°N to 90°S and 90°N.

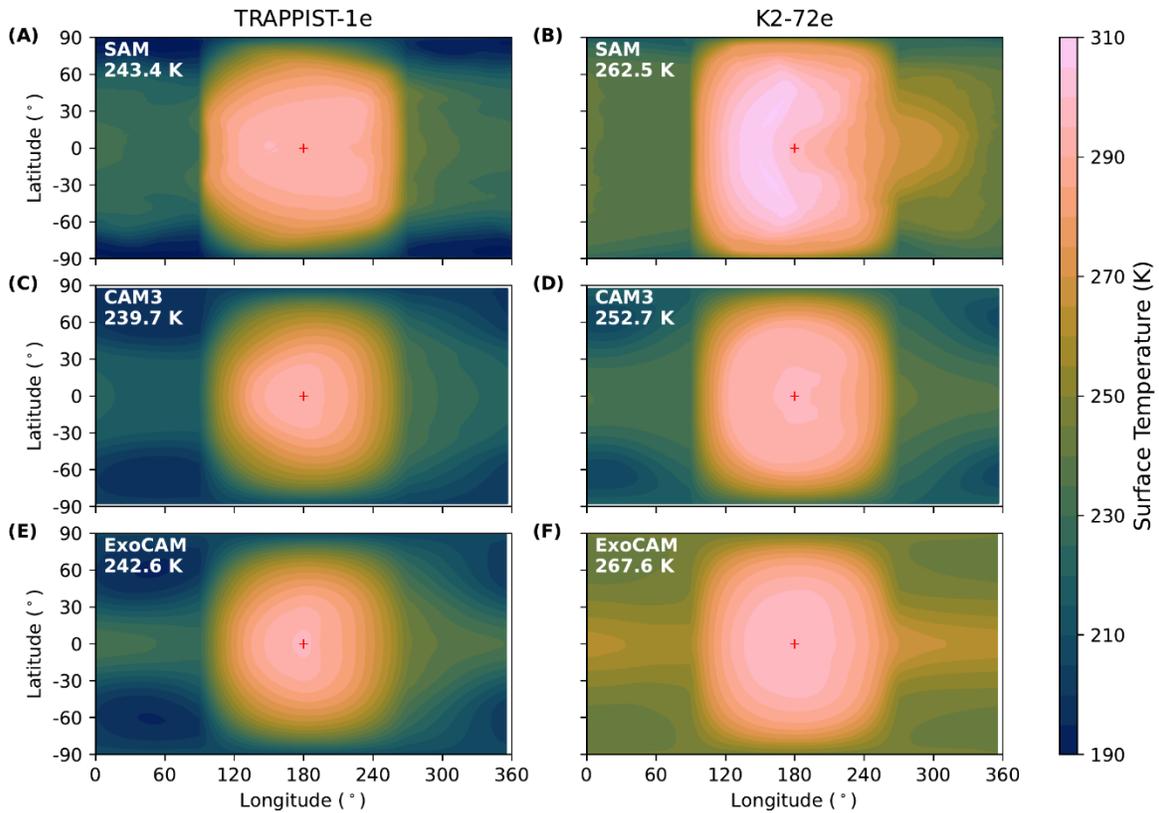

**Figure 8. Surface temperatures simulated by the cloud-permitting model SAM and the two general circulation models CAM3 and ExoCAM.** Left panels: planet TRAPPIST-1e, and right panels: planet K2-72e. (A) & (B) are for SAM, (C) & (D) are for CAM3, and (E) & (F) are for ExoCAM. The red cross marks the substellar point. The global-mean value is listed at the top left corner of each panel. For SAM, the grid spacing is 4 km by 4 km. In this and following figures, the values of SAM are the mean of 10 Earth days (unless mentioned otherwise), and the values of CAM3 and ExoCAM are for the mean of 5 Earth years.

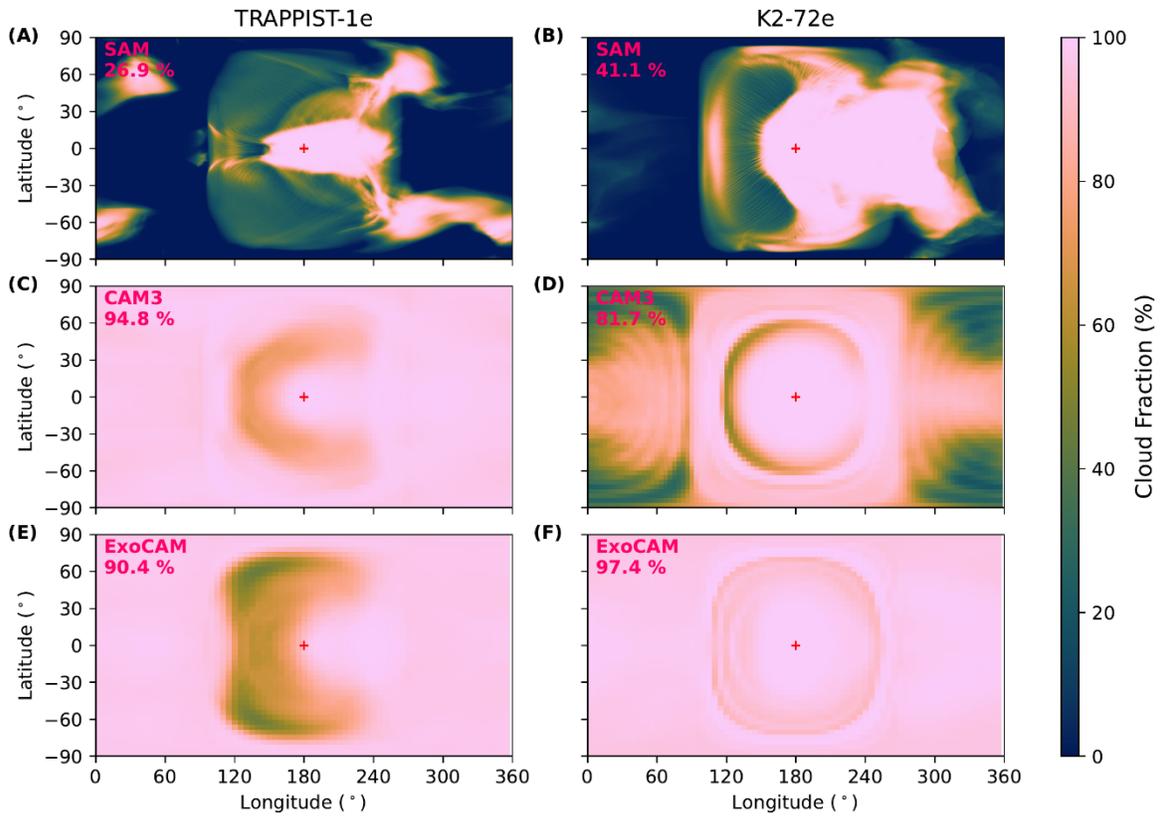

**Figure 9. Same as the supplementary Figure 8 above, but for cloud fraction.** Cloud fraction in SAM is much less than those in CAM3 and ExoCAM especially at the nightside and at the high latitudes of the dayside. The red cross marks the substellar point. SAM and CAM3/ExoCAM use different definitions for cloud fraction, please see Text 3 above.

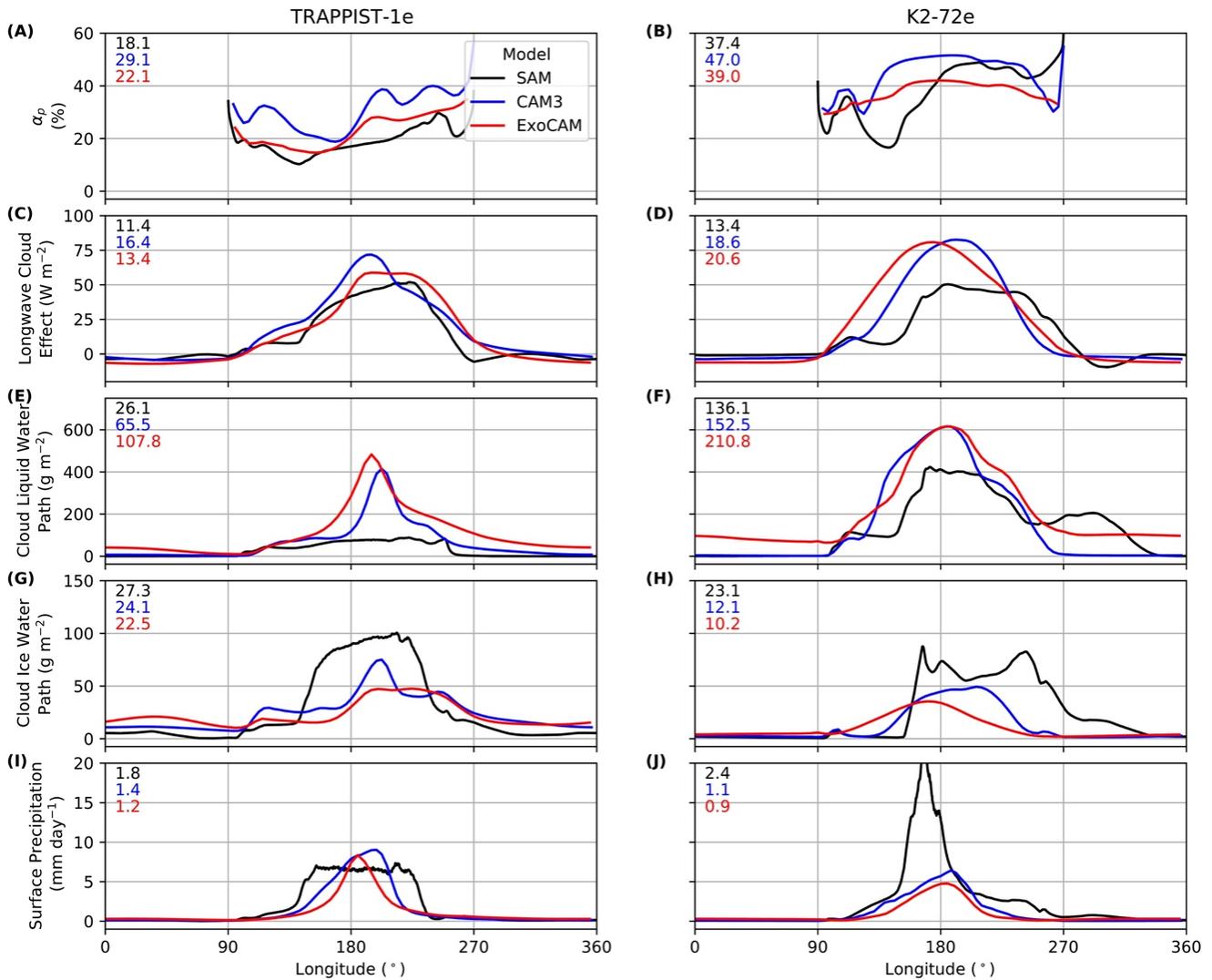

**Figure 10**. **Model intercomparison in clouds and precipitation between SAM, CAM3, and ExoCAM.** Left panels are for TRAPPIST-1e, and right panels are for K2-72e. (A) & (B) are for planetary albedo (%), (C) & (D) are for longwave cloud radiative effect (W m$^{-2}$), (E) & (F) are vertically-integrated cloud liquid water amount (g m$^{-2}$), (G) & (H) are for cloud ice water amount (g m$^{-2}$), and (I) & (J) are for surface precipitation rate (mm day$^{-1}$). Black color is for SAM, blue is for CAM3, and red is for ExoCAM. All these variables are for time- and meridional-mean. Global-mean values are listed with corresponding colors in the top left corner of each panel. The planetary albedo is undefined on the nightside.

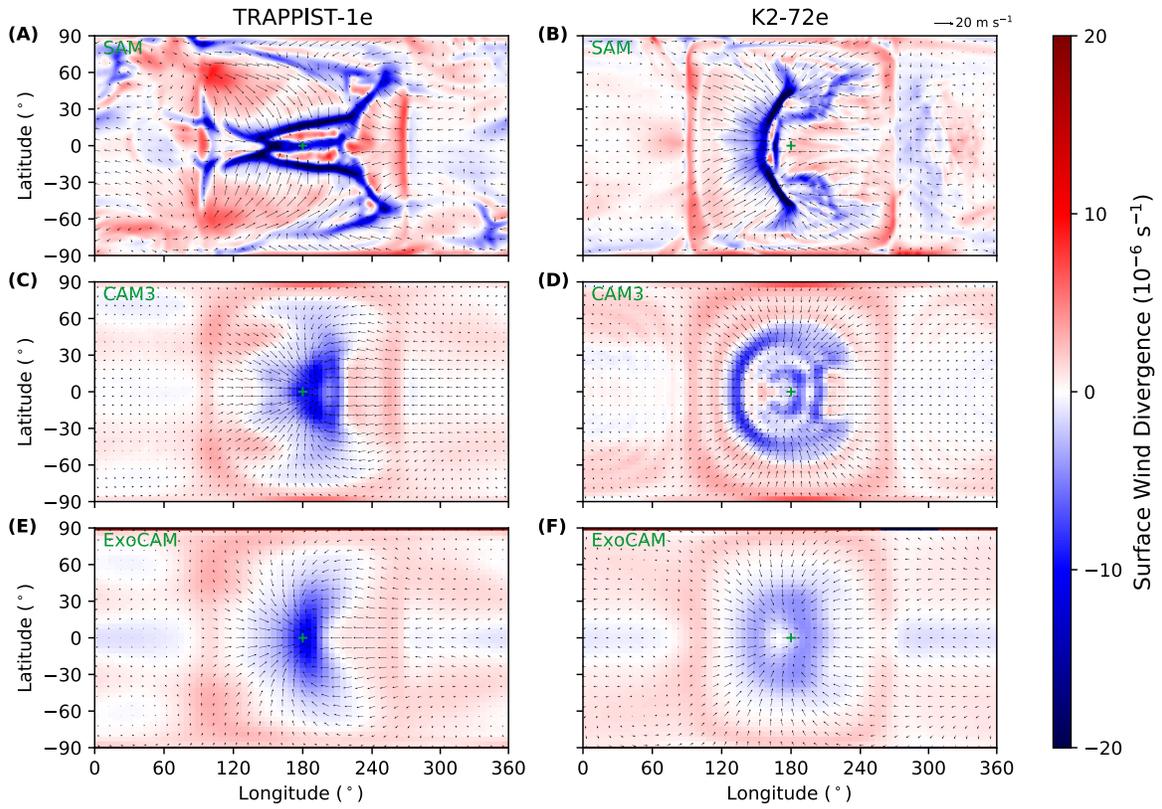

**Figure 11**. **Model intercomparison in near-surface horizontal winds (vectors) and divergence of the winds (color shading, negative is convergence and positive is divergence) between SAM, CAM3, and ExoCAM.** Left panels are for TRAPPIST-1e, and right panels are for K2-72e. The green cross marks the substellar point.

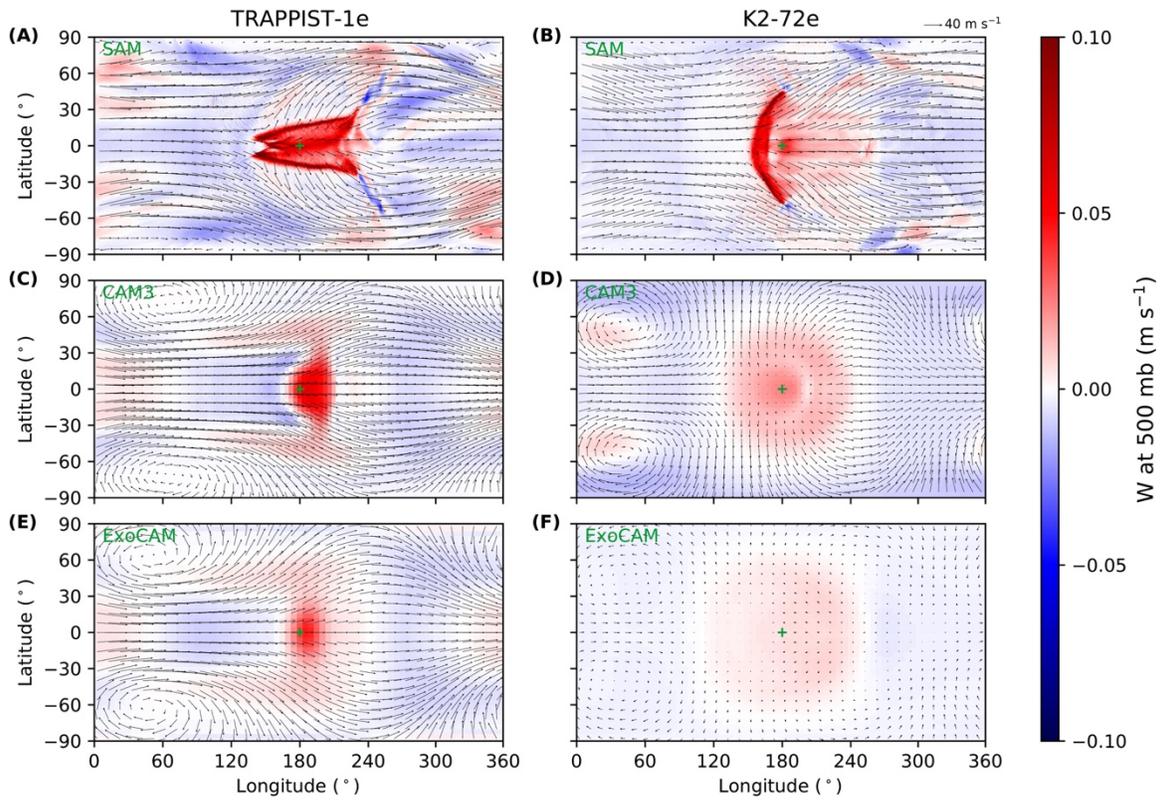

**Figure 12**. **Model intercomparison in large-scale atmospheric circulation between SAM, CAM3, and ExoCAM.** Left panels are for TRAPPIST-1e, and right panels are for K2-72e. The color shading is for vertical velocity at the level of 500 hPa, and the vectors are for the horizontal winds at the level of 200 hPa. The green cross marks the substellar point.

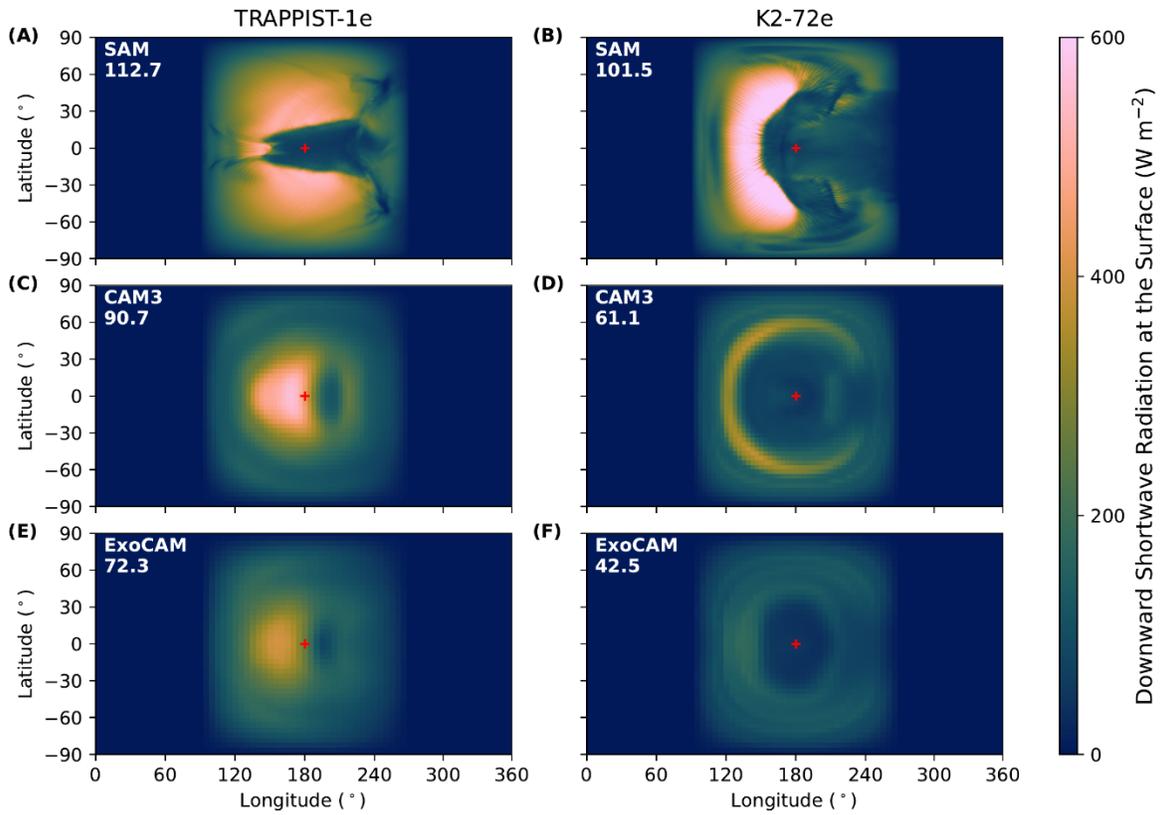

**Figure 13**. **Same as the supplementary Figure 8, but for downward shortwave radiation flux at the surface.** Left panels are for TRAPPIST-1e, and right panels are for K2-72e. The red cross marks the substellar point. The global-mean value in units of W m$^{-2}$ is listed in the top left corner of each panel.

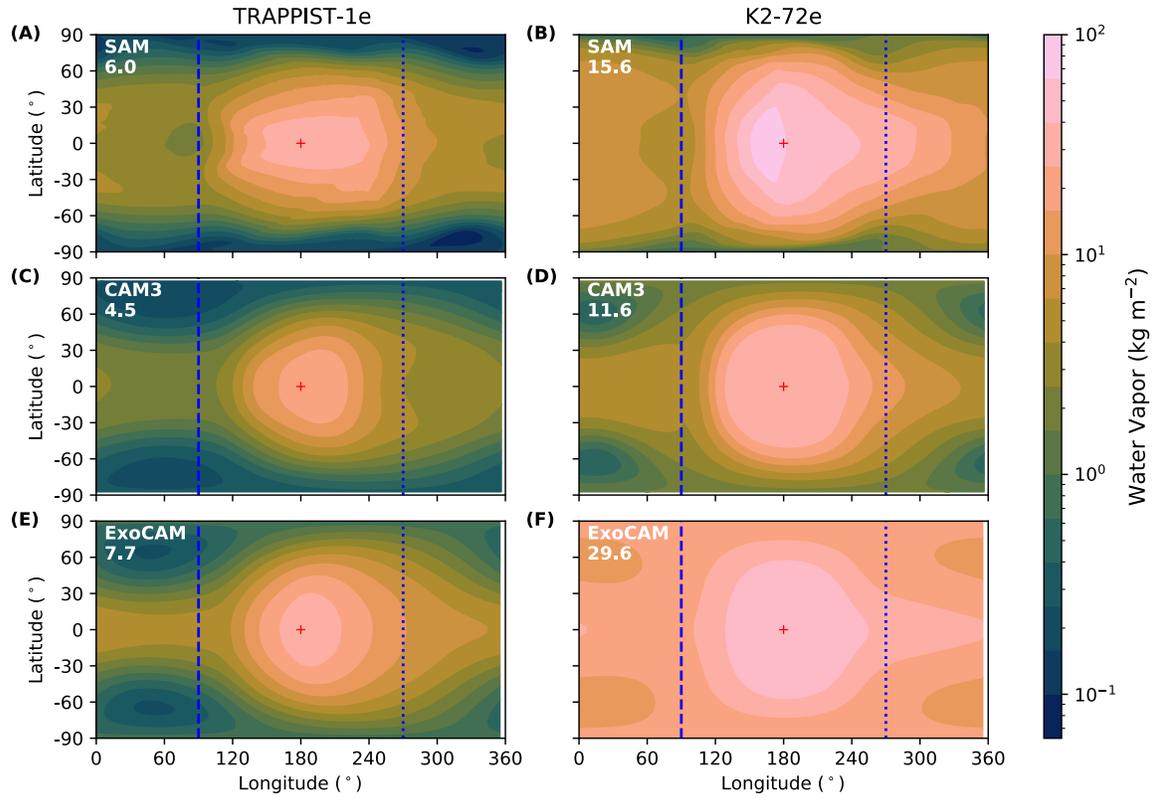

**Figure 14**. **Same as the supplementary Figure 8, but for vertically-integrated water vapor concentration.** Left panels are for TRAPPIST-1e, and right panels are for K2-72e. The red cross marks the substellar point. The global-mean value in units of kg m$^{-2}$ is listed in the top left corner of each panel.

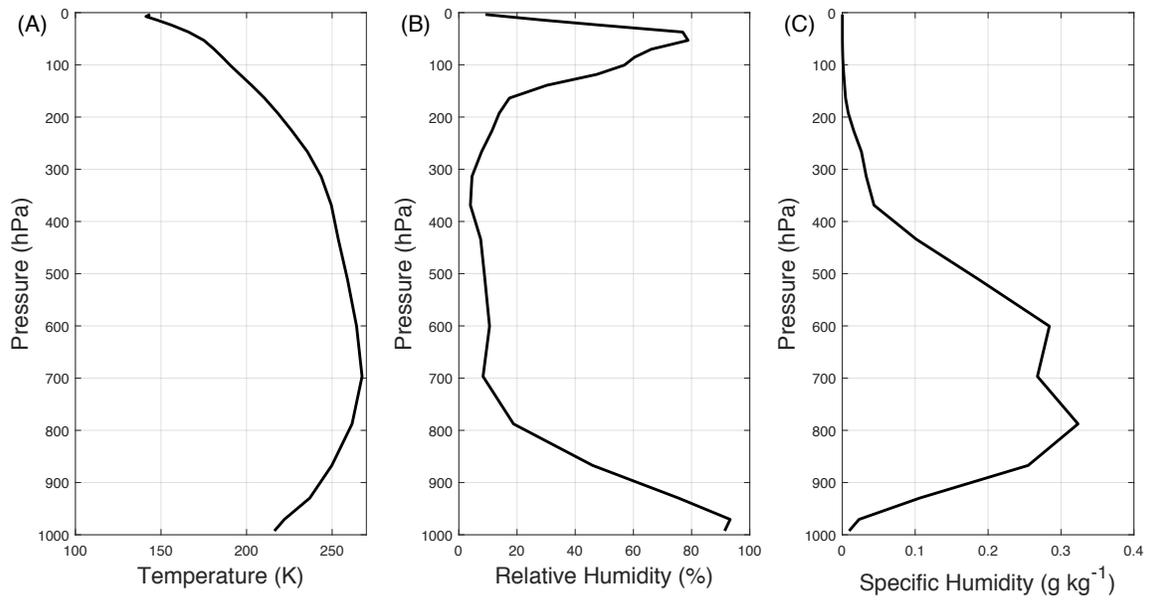

**Figure 15**. **Profiles of air temperature (A), relative humidity (B), and specific humidity (C) over the anti-stellar point in one CAM3 experiment.** The stellar flux at the substellar point is 1400 W m$^{-2}$, the rotation period (= orbital period) is 22.12 Earth days, and the star temperature is 3300 K.

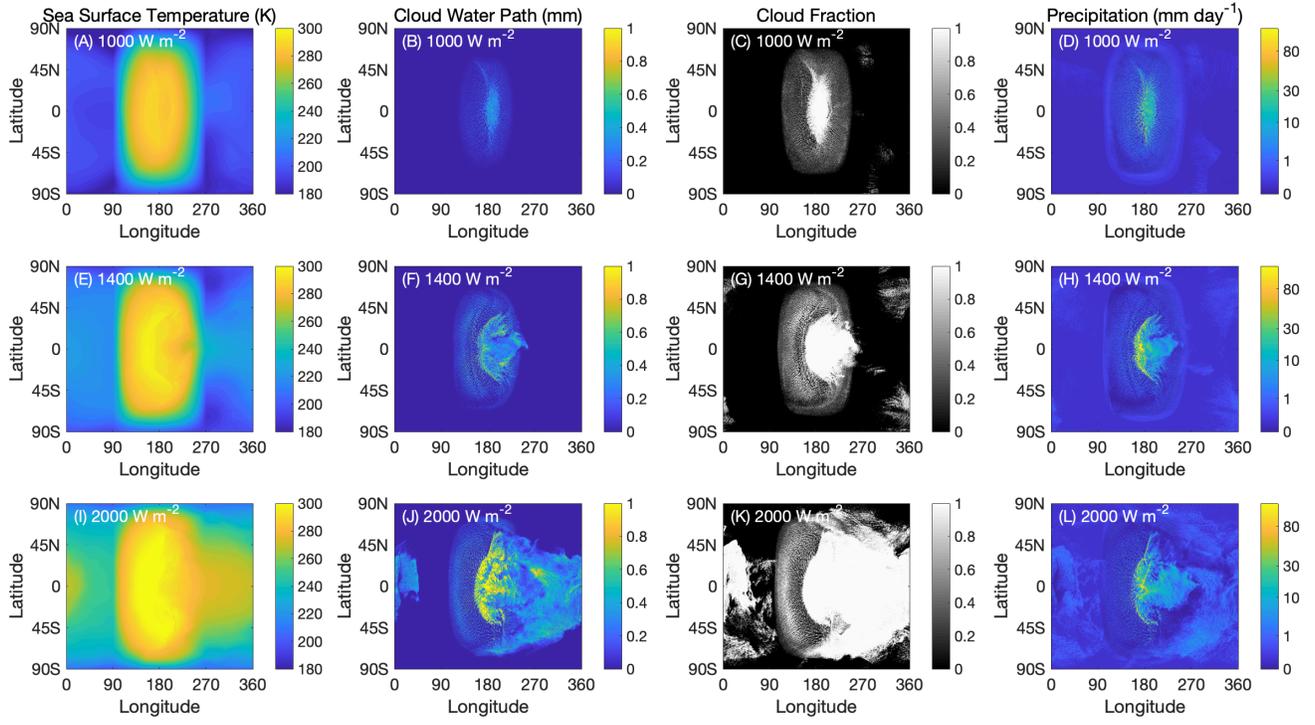

**Figure 16**. **Snapshots of sea surface temperature (first column), vertically-integrated cloud water path (second column), total cloud fraction (0-1, third column), and surface precipitation (fourth column) in the stabilizing cloud feedback experiments of SAM.** The stellar fluxes are 1000, 1400, and 2000 W m$^{-2}$ for the upper, middle, and bottom rows, respectively. The substellar point is at the center of each panel.

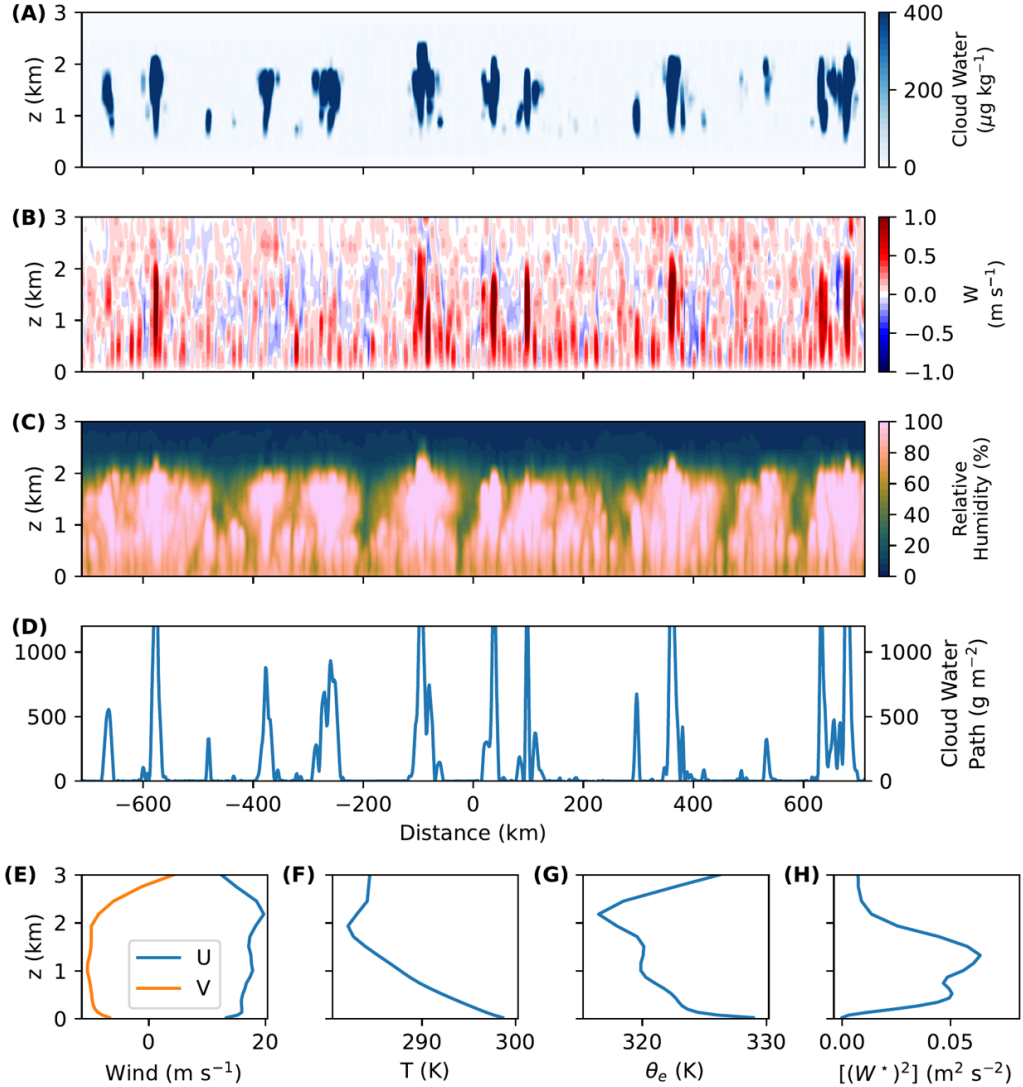

**Figure 17. The formation of cloud streets**. (A) cloud water concentration ($\mu$g kg$^{-1}$); (B) vertical velocity (m s$^{-1}$); (C) relative humidity (%); (D) vertically-integrated cloud water (g m$^{-2}$); (E) winds in zonal and meridional directions (m s$^{-1}$); (F) temperature profile (K); (G) equivalent potential temperature profile (K); and (H) the square of the vertical velocity anomalies: $[(W^{*2})]$ where $W$ is the vertical velocity, the star represents the deviation from the domain mean, and the square brackets denote the domain mean. The selected region is along the dashed yellow line in panel B (the K2-72e experiment) of Figure 5 in the main text. The wind shear is strong in the levels between 2 and 3 km, but it is weak in the convective region between near surface and 2 km (E), because convection is effective in vertically re-distributing the momentum. The air in the boundary layer is in convection as characterized by the surface being warmer than the adjacent overlying air (F & G) and by the weak equivalent potential temperature gradient between ~1 and 2 km (G). The vertical mixing is strong below 2 km and becomes much weaker above 2 km (H).

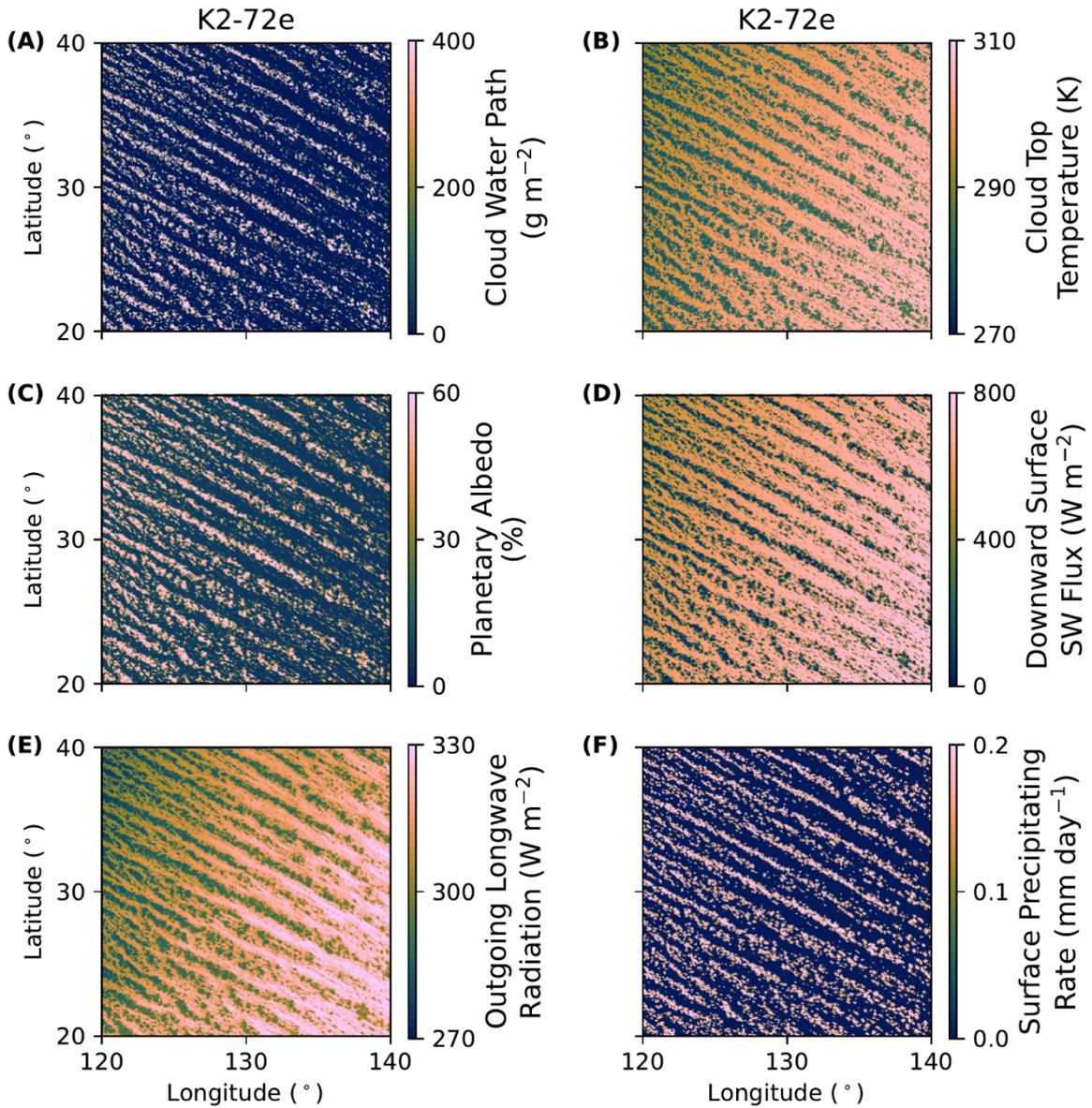

**Figure 18. The climatic effects of cloud streets.** (A) vertically-integrated cloud liquid and ice water mass (g m$^{-2}$), (B) cloud top temperature (K), (C) planetary albedo (%), (D) downward shortwave flux at the surface (W m$^{-2}$), (E) outgoing longwave radiation (OLR) to space (W m$^{-2}$), and (F) surface precipitating rate (mm day$^{-1}$). The selected region is the same as the panel B (the K2-72e experiment) of Figure 5 in the main text.

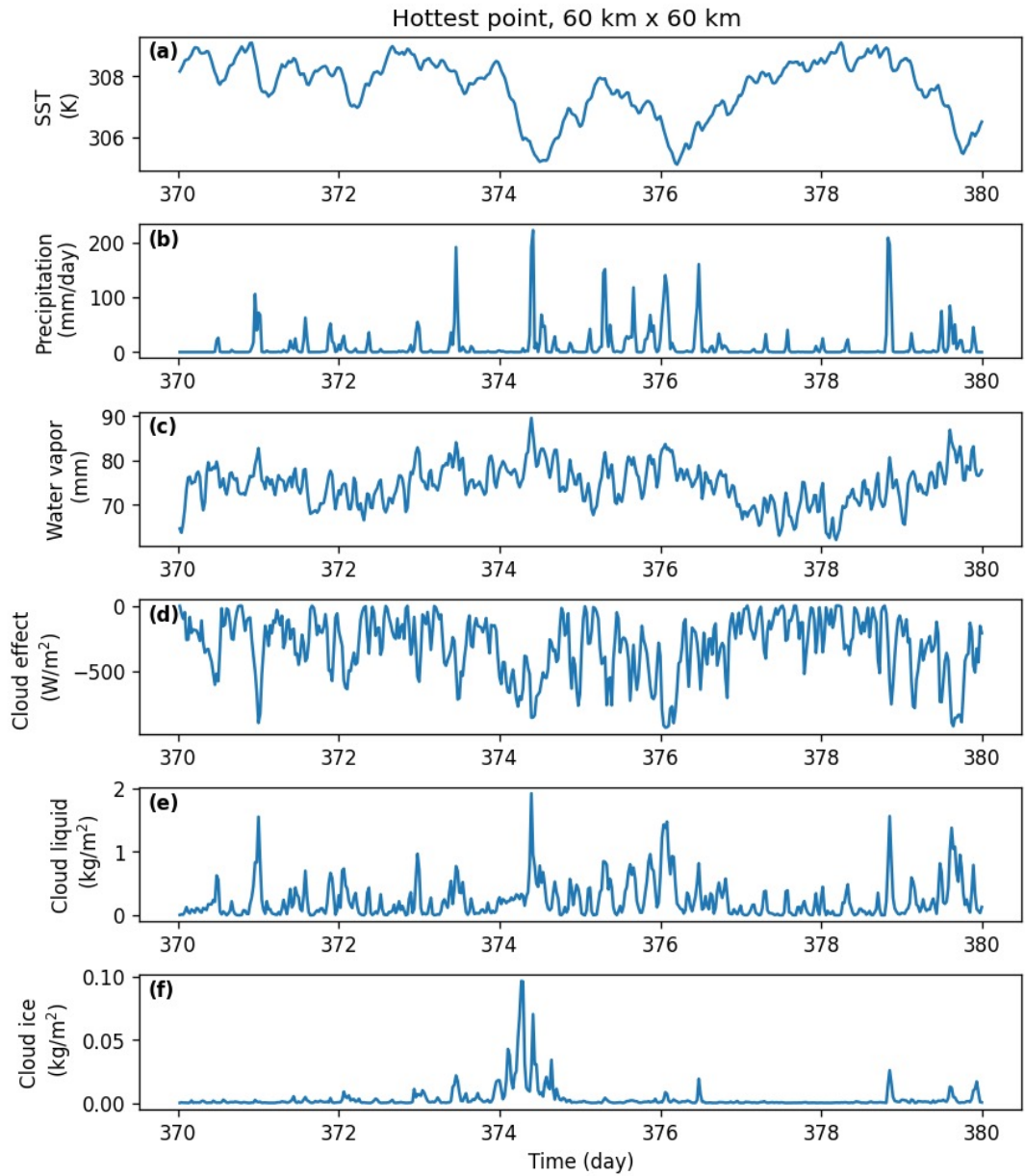

**Figure 19**. **Time series of surface temperature (a), precipitation (b), water vapor concentration (c), net cloud radiative effect at the top of the model (d), vertically-integrated cloud liquid water amount (e, kg/m$^2$), and vertically-integrated cloud ice water amount (f, kg/m$^2$), at a small domain region (60 km × 60 km) on the east side of the substellar point in one of SAM's quasi-global experiments**. This region has the highest surface temperature, and it locates on the equator and 36º in longitude away from the substellar point. The stellar flux at the substeller point is 2000 W m$^{-2}$, the rotation period (= orbital period) is 16.93 Earth days, and the horizontal resolution is 20 km by 20 km. The model is coupled to a slab ocean with a uniform depth of 1 m. These variables are the mean values over a small domain (60 km × 60 km), following the Extended Data Figure 6 in Seeley & Wordsworth (2021).

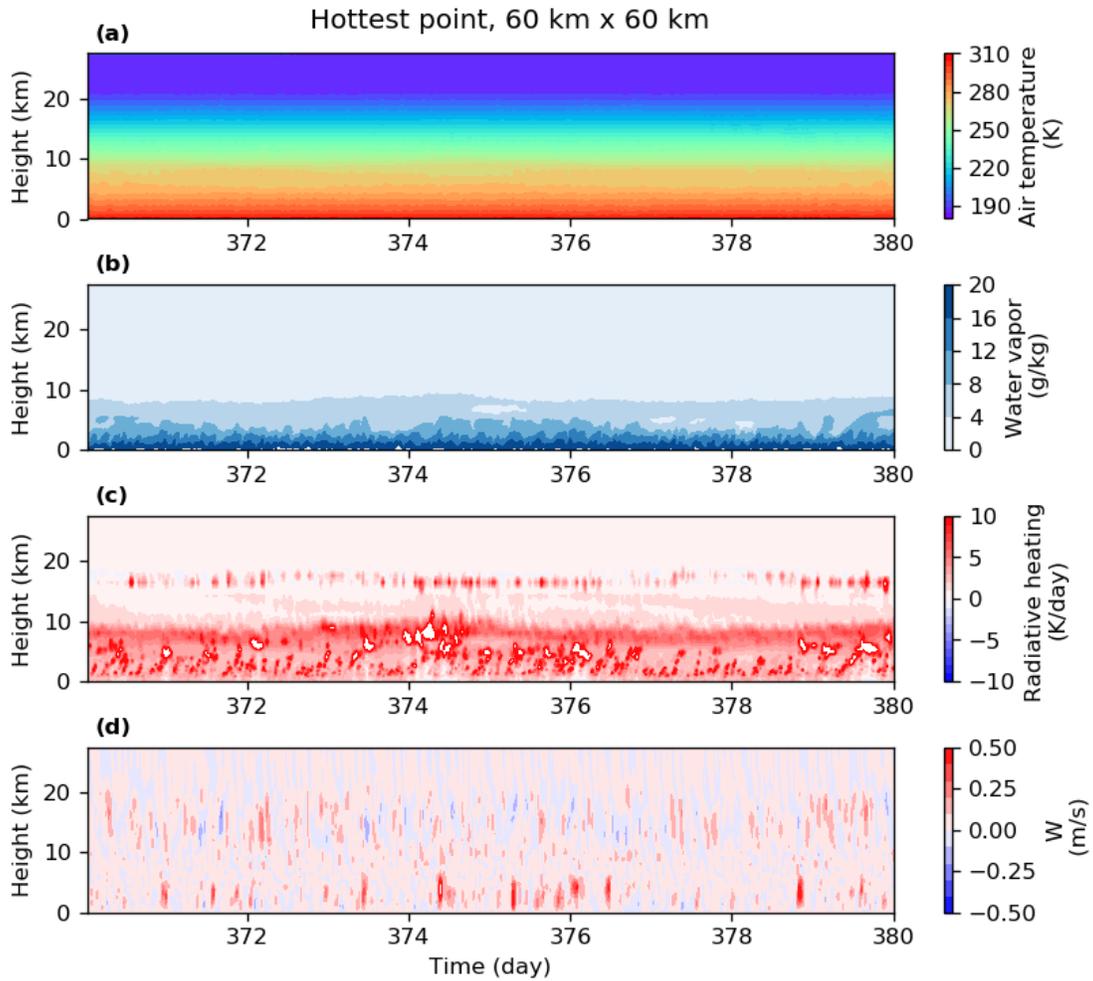

**Figure 20**. **Same as the Supplementary Figure 19 above, but for air temperature, water vapor concentration, net radiative (shortwave plus longwave) heating rate, and vertical velocity as a function of time and height, over the hottest region (60 km × 60 km) in the quasi-global simulation of SAM.**

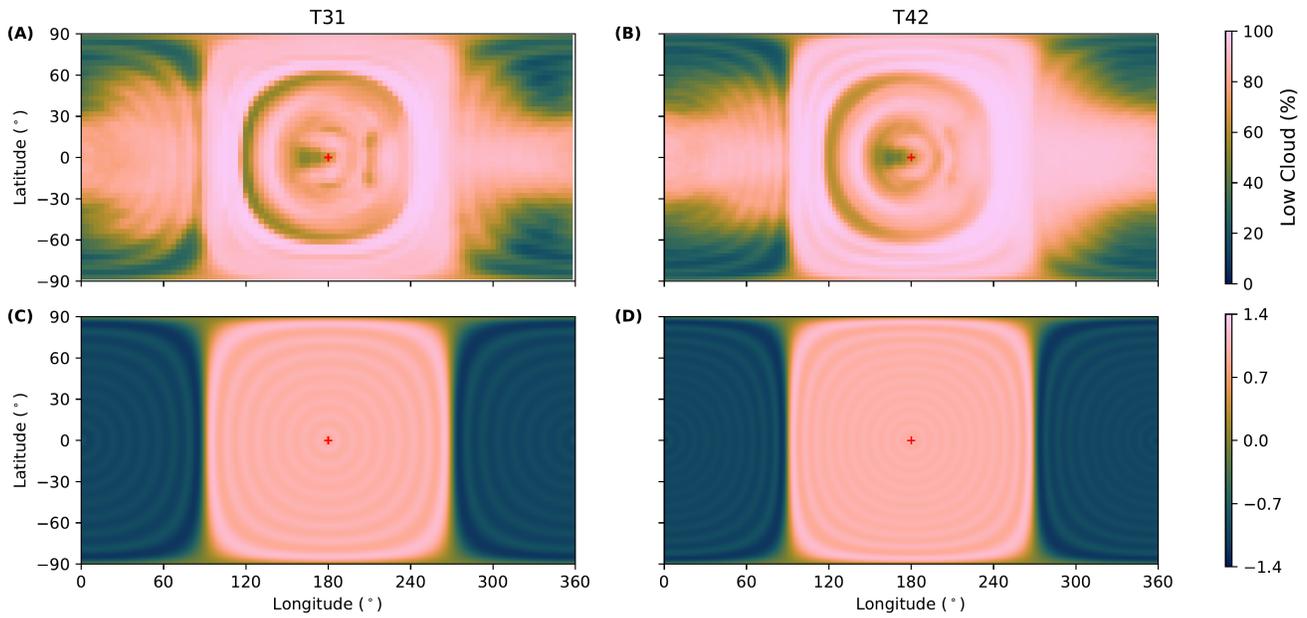

**Figure 21. The simulated cloud fraction and the truncated field**. Left panels: the resolution is T31, ~3.75° by 3.75°. Right panels: the resolution is T42, ~2.8° by 2.8°. The upper panels are the simulated low-level cloud fractions by CAM3, and the lower panels are truncated fields (see Text 9 above).

**Video 1** (separate file online). Time variability of the vertically-integrated cloud liquid and ice water amount (g m$^{-2}$) in the simulation of TRAPPIST-1e with a horizontal resolution of 4 km. The time interval of each frame is 1 hour.

**Video 2** (separate file online). Time variability of the vertically-integrated cloud liquid and ice water amount (g m$^{-2}$) in the simulation of K2-72e. This movie is the same as the supplementary Video 1, but for the planet K2-72e.

**Video 3** (separate file online). Time variability of the vertically-integrated cloud liquid and ice water amount (g m$^{-2}$) in the simulation of TRAPPIST-1e. This movie is the same as the supplementary Video 1, but the simulation uses a higher resolution, 2 km.